\documentclass[a4paper,fleqn,usenatbib,useAMS]{mnras}

\usepackage{graphicx}	
\usepackage{amsmath}	
\usepackage{amssymb}	
\usepackage{multicol}        
\usepackage{bm}		
\usepackage{pdflscape}	

\usepackage[T1]{fontenc}
\usepackage{ae,aecompl}

\usepackage{newtxtext,newtxmath}

\usepackage{etoolbox}
\makeatletter
\patchcmd\@combinedblfloats{\box\@outputbox}{\unvbox\@outputbox}{}{\errmessage{\noexpand patch failed}}
\makeatother

\title[Transit Multiplicity in Planet Occurrence Rates]{Accounting for Incompleteness due to Transit Multiplicity in \emph{Kepler} Planet Occurrence Rates}

\author[Zink, Christiansen, and Hansen]{Jon K. Zink$^{1}$ \thanks{E-mail: \href{mailto:jzink@astro.ucla.edu}{jzink@astro.ucla.edu}}, Jessie L. Christiansen$^{2}$, and Bradley M. S. Hansen$^{1}$
\\
$^{1}$Mani L. Bhaumik Institute for Theoretical Physics, Department of Physics and Astronomy, University of California, Los Angeles, CA 90095
\\
$^{2}$NASA Exoplanet Science Institute, California Institute of Technology, Pasadena, CA 91106}

\date{Last updated 2018 December 18}

\pubyear{2018}

\begin{document}
\label{firstpage}
\pagerange{\pageref{firstpage}--\pageref{fig:post}}
\maketitle

\begin{abstract}
We investigate the role that planet detection order plays in the \emph{Kepler} planet detection pipeline. The \emph{Kepler} pipeline typically detects planets in order of descending signal strength (MES). We find that the detectability of transits experiences an additional 5.5\% and 15.9\% efficiency loss, for periods $<200$ days and $>200$ days respectively, when detected after the strongest signal transit in a multiple-planet system. We provide a method for determining the transit probability for multiple-planet systems by marginalizing over the empirical \emph{Kepler} dataset. Furthermore, because detection efficiency appears to be a function of detection order, we discuss the sorting statistics that affect the radius and period distributions of each detection order. Our occurrence rate dataset includes radius measurement updates from the California Kepler Survey (CKS), \emph{Gaia} DR2, and asteroseismology. Our population model is consistent with the results of \citet{bur15}, but now includes an improved estimate of the multiplicity distribution. From our obtained model parameters, we find that only $4.0\pm4.6\%$ of solar-like GK dwarfs harbor one planet. This excess is smaller than prior studies and can be well modeled with a modified Poisson distribution, suggesting that the \emph{Kepler} Dichotomy can be accounted for by including the effects of multiplicity on detection efficiency. Using our modified Poisson model we expect the average number of planets is $5.86 \pm 0.18$ planets per GK dwarf within the radius and period parameter space of \emph{Kepler}. 
\end{abstract}


\section{Introduction} 

The \emph{Kepler} mission has revolutionized our understanding of the frequencies and properties of planets around Sun-like stars. With the final data release DR25, providing all of the data up until the failure of two reaction wheels \citep{mat17}, the primary phase of the project has officially concluded. Within this span, \emph{Kepler} has provided evidence for $\approx4,500$ transiting exoplanets.\footnote{\url{https://exoplanetarchive.ipac.caltech.edu}} Nearly 50\% of these candidates have been confirmed or validated \citep{row14,mort16}, demonstrating that planets are common and widespread in the Milky Way.\

There have been many attempts to quantify the frequency of planetary systems and the properties (radius and orbital period) of the planets themselves \citep{bor11,catshao,you11,how12,bat13,fress13,pet13a,dz13,dre13,mul15,dre15,bur15,muld15,sil15}, with a special attention given to attempting to characterize the frequency of planets with Earth-like properties.  One of the most challenging aspects of estimating these occurrence rates is understanding the completeness of the known exoplanet sample. The automation provided by the \emph{Kepler} pipeline has produced a systematic method of detecting transiting exoplanets and thus offers the prospect of a rigorous determination of the survey completeness. With ~3.5 years of nearly continuous light curves of ~200,000 stars, it is possible to investigate period ranges out to 500 days. Furthermore, the high precision of the \emph{Kepler} light detector has permitted the discovery of planets with radii $r<1 r_{\earth}$.\

Since the completion of the \emph{Kepler} survey, several studies have used this data set to extract population parameters. \citet{pet13a}, using their own \emph{TERRA} pipeline, implemented an \emph{Inverse Detection Method}, where the population CDF (Cumulative Distribution Function) is divided by the detection efficiency. This study also introduced the idea of synthetic planet injections into the \emph{Kepler} light curves to map completeness. Here, artificial transits were injected into the \emph{Kepler} light curves, and the recovery fraction in the \emph{TERRA} pipeline was used to understand the \emph{Kepler} detection efficiency. To avoid confusion from multiple planet transits the \citet{pet13a} occurrence rate calculation only included the highest SNR (Signal to Noise Ratio) planet in each system, ignoring any multiplicity. To characterize the official \emph{Kepler} completeness, \citet{chr15} performed a pixel-level transit injection test to empirically measure how well the  pipeline would detect various types of planets. This is discussed in more detail in Section \ref{sec:inject}. The results of this study were then used by \citet{bur15} to perform a \emph{Poisson Process Analysis}, where a Bayesian framework is implemented to determine the best population model parameters. The current work employs a similar method. \

Planet multiplicity introduces detection biases above and beyond those to which single transit systems are subject. When faced with a system of multiple transiting planets, the \emph{Kepler} pipeline will typically find the largest MES (Multiple Event Statistic; comparable to SNR) signal, fit the transit function, and then discard the corresponding data points. The width of discarded data is $3\times$ the transit duration, with $1.5\times$ removed on each side of the transit center. Very few TTVs (Transit-Timing Variations) are large enough to escape this window. Such deletion is necessary to avoid confusion when looking for additional planets, but introduces data gaps into the light curve as noted by \citet{sch17}. These gaps becomes more invasive in higher multiplicity systems where significant data is being discarded. With each planet removed, the available data set shrinks. This effect creates ``swiss cheese''-like holes in the light curves, where the number of holes increases with each detected planet. Beyond possible gaps in the light curve, the \emph{Kepler} pipeline fails to detect some short-period planets because of a harmonic fitting function \citep{chr13}. Here the pipeline attempts to remove sinusoidal variations in the light curve caused by stellar activity, but in doing so, the procedure can overfit a true planet signal and make low SNR planets difficult to detect. To clarify, the baseline wobble from the dataset is removed using a spline smoothing function. The harmonic fitting function is specifically looking for sinusoidal variations in the light curve. This function may or may not be applied, depending on whether the pipeline is able to detect such periodic variation in the light curve. In multiple-planet systems, the harmonic fitter can also overfit the periodic variations caused by transits and remove true signals. Because the pipeline follows these procedures, the order of planet detection can affect it's detectability.\

Our goal in this paper is to assess the effect of planet multiplicity and detection order on the completeness of the Kepler results. In Section \ref{sec:stell} and \ref{sec:plan} we describe our methods of stellar and planet selection. In Section \ref{sec:inject} we show that detection order affects the detection efficiency for a given planet. In Section \ref{sec:incl} we describe how we account for mutual inclination within this study. In Section \ref{sec:detc}, we lay out our process of accounting for overall detection efficiency. In Section \ref{sec:like} we present our expanded likelihood function used to calculate the posterior for the population parameters. In Section \ref{sec:discuss}, we discuss the results of our fitting method and the implications of our multiplicity parameters. We provide concluding remarks in Section \ref{sec:conc}.\

\section{ Stellar Selection}
\label{sec:stell}
Using the final release of \emph{Kepler} data (DR25) which includes Q1-Q17, we select a stellar sample for use in creating a detection efficiency map that accounts for \emph{Kepler} completeness. We use the stellar parameters provided by \citet{mat17} with improved radius values derived from \emph{Gaia} DR2 \citep{ber18}. The updates from \emph{Gaia} DR2 have yet to provide updated corresponding mass values. Thus we must still utilize the \emph{Kepler} DR25 stellar mass parameters (200,038 stars in total). To focus on the occurrence of planets around solar-like GK dwarfs, we only include stars with $T_{\rm eff}>4200,K$ and $T_{\rm eff}<6100K$ (135,494 stars remain). It is also important for completeness mapping that each star has a stellar radius and mass measurement available. ``Null'' values for either of these fields result in omission (133,056 stars remain). To avoid the inclusion of giants we limit the sample to $log(g)\ge4$ and $R_{\star}\le 2R_{\sun}$ (96,167 stars remain). We also place requirements on the duty cycle ($f_{duty}$) and the time length of the light curve ($data_{span}$). These are $f_{duty}>0.6$ and $data_{span}>2$ years are made (86,679 stars remain). The $f_{duty}$ limit requires that 60\% of $data_{span}$ has been collected. This ensures that a significant portion of the light curve is filled, while still including stars lost in the Q4 CCD loss \citep{bat13}. Time-varying noise measurements have been provided in the DR25 dataset through a value known as CDPP (Combined Differential Photometric Precision; \citealt{chr12}). This parameter has been calculated for every field star over 14 different time periods: 1.5, 2.0, 2.5, 3.0, 3.5, 4.5, 5.0, 6.0, 7.5, 9.0, 10.5, 12.0, 12.5, and 15.0 hours \citep{mat17}. These values correspond to the amount of noise a planet signal will need to exceed, given a transit duration, to generate a $1\sigma$ detection. By requiring stars to have a $\rm CDPP_{7.5h}<1000$ ppm, we minimize the inclusion of stellar and instrumental fluctuations (74 stars exceed this limit). From this we produce a stellar sample of 86,605 solar-like stars.\

\setcounter{figure}{0}
\begin{figure*}
\hfill \includegraphics[width=8.5cm]{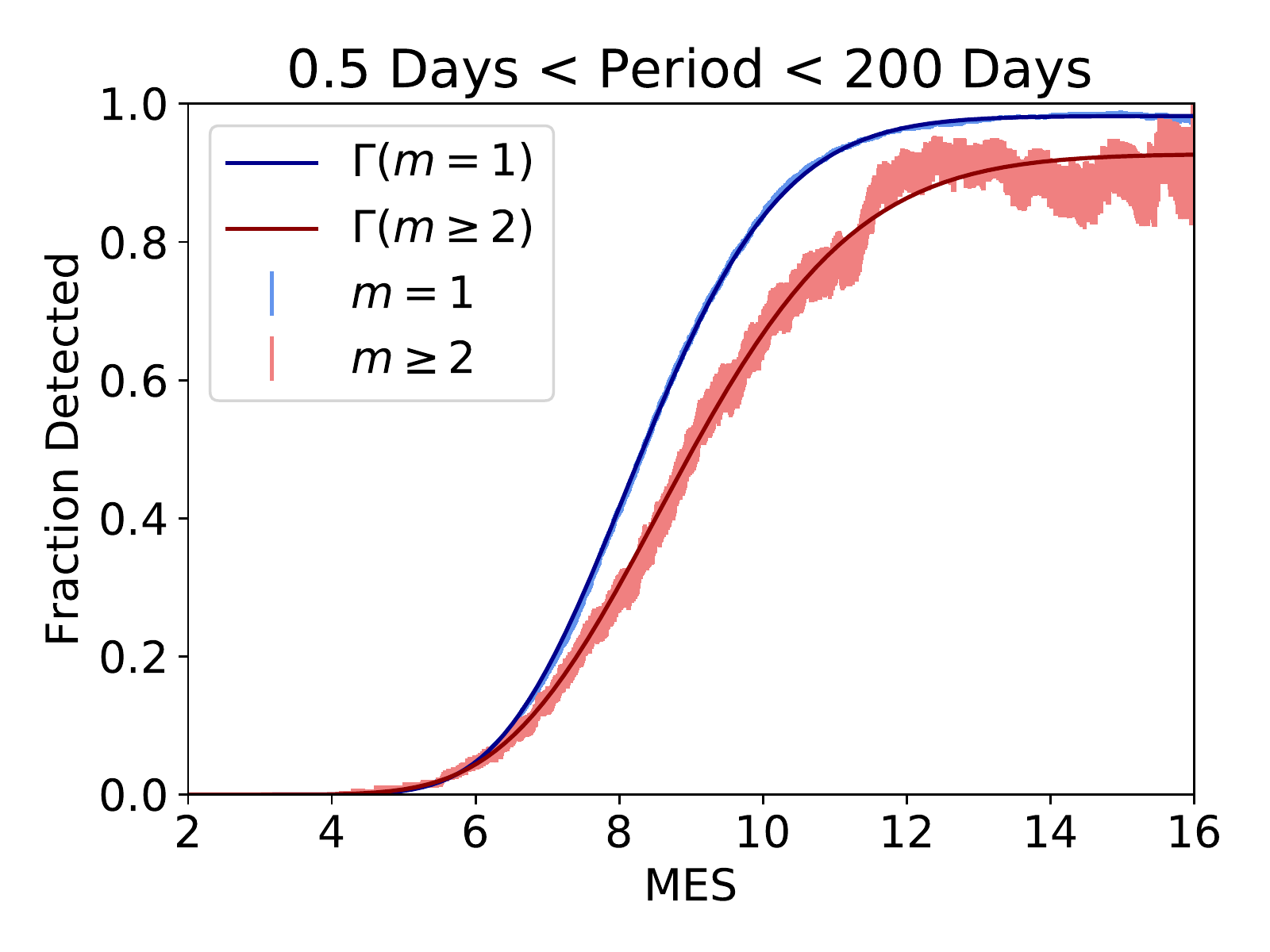} \hfill \includegraphics[width=8.5cm]{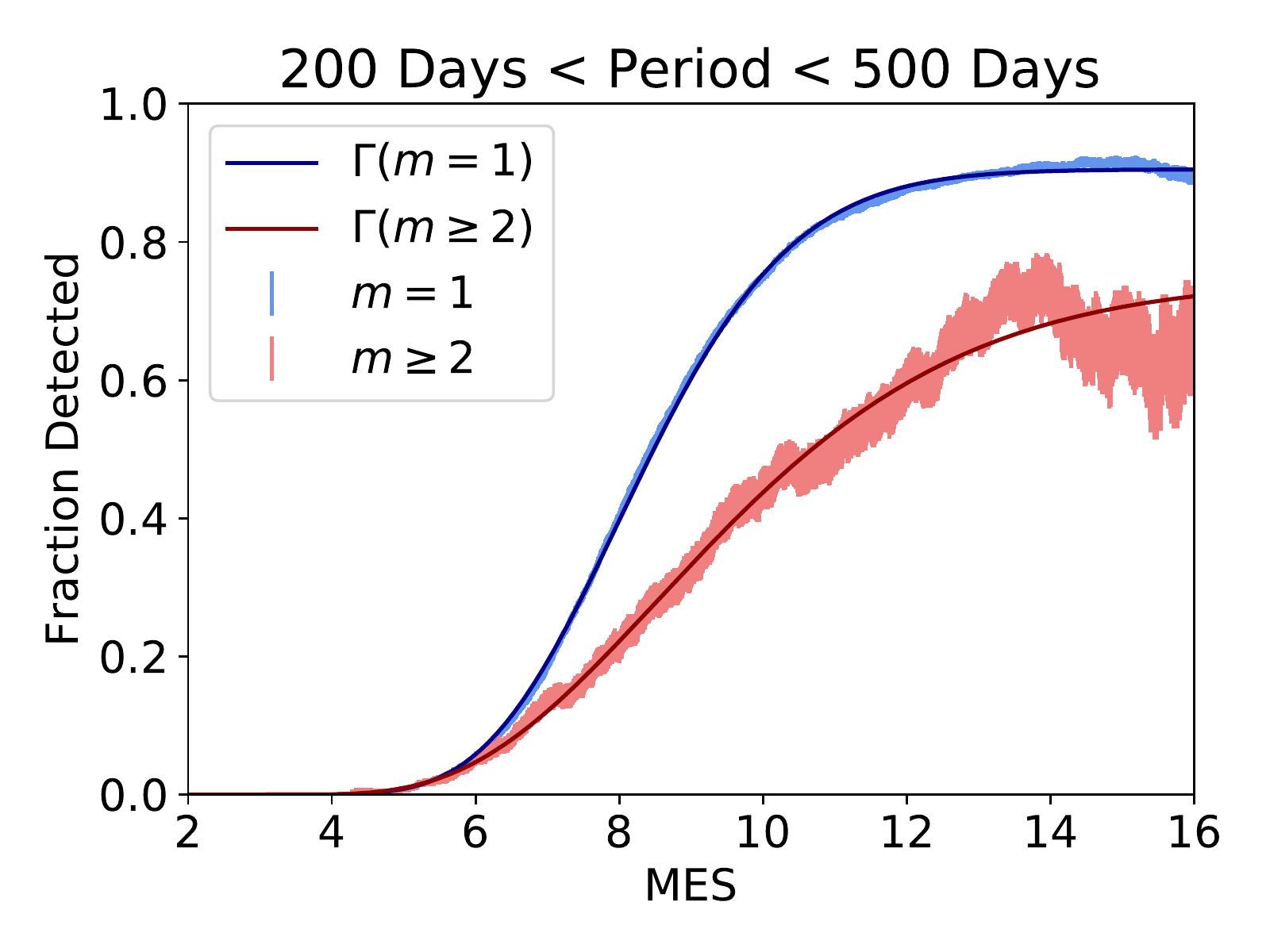} \hfill
\caption{The smoothed recovery fraction at each MES bin. The vertical lines (light blue and red) represent the uncertainty in each bin under the assumption of a binary distribution. The bin values are plotted at the center of each bin. The solid lines (dark blue and red) represent the $\Gamma_{CDF}$ distribution fit. The parameters of this model were fit using a $\chi^2$ minimization.\label{fig:mes}}
\end{figure*}

\section{Planet Selection}
\label{sec:plan}
When available we utilize the updated planetary parameters provided by the California \emph{Kepler} Survey (CKS) \citep{pet17,joh17} and the asteroseismic updates provided by \citet{vanE18a}. One of the main advantages for the inclusion of these updates is the improved planet radius measurements. Since our study, like others, does not account for parameter uncertainty, such improvements are essential for accurate occurrence rates. Where CKS and asteroseismic data are unavailable, the measurements provided by the \emph{Kepler} DR25 catalog \citep{tho18}, in conjunction with the \emph{Gaia} DR2 radius updates \citep{ber18}, are implemented. Through private communication, it was indicated that this early release of \emph{Gaia} data may contain some planet radius outliers. To combat this issue, we test the radius values against the \emph{Kepler} DR25 catalog. When the updated \emph{Gaia} measurements differ from the \emph{Kepler} DR25 data by $>3\sigma$, we utilize the \emph{Kepler} DR25 radius measurements. Overall, 19 planets exceed this outlier limit (statistically, we would expect only 8). All period measurements are drawn from the light curves; thus, improved measurements from \emph{Gaia} and CKS have no effect on the inferred period measurements. We use the periods provided in the \emph{Kepler} DR25 catalogs. Both the CKS and \emph{Kepler} DR25 provide flags for false positives. We include data from both CONFIRMED and CANDIDATE planets in DR25 and $CKS_{fp} =FALSE$ in the CKS update. To further avoid contamination from false positives, we only include planets with periods $.5<p<500$ days and radii $.5<r<16 r_{\earth}$. Periods beyond 500 days have been noted to be highly contaminated by false positives because they barely meet the three transit limit of the pipeline \citep{mul15}.\

Our period and radii range exceeds the conservative cutoffs adopted by many previous studies, but is necessary when exploring the effects of multiplicity. Often planetary systems span the entire range of the \emph{Kepler} parameter space, thus the inclusion of nearly all the planets is needed for an accurate calculation. There exist 3 multi-planet systems (KIC: 3231341, 11122894, and 11709124) where one planet within the system fall beyond the range of this study. We only select the planets from these system that lie within our radius and period cuts. The inclusion of these planets is useful in providing a stronger statistical argument. Although some of the known planets, in these 3 systems, extend beyond the bounds of this study, we expect many other systems within the dataset to contain planets beyond the range of our selection bounds. Furthermore, if we include the planets that lay beyond our radius and period cuts, our analysis we will artificially inflate the number of inferred planets within this range.\

The accuracy of the \emph{Kepler} detection order (``TCE Planet Number'') can be affected by systems with existing false positives. When removing these data points, we manually ensure that the detection order only reflects the order in which valid KOIs (\emph{Kepler} Objects of Interest) are detected. For example, a system with 5 ``real'' KOIs and 1 false positive would have detection orders ranging from 1-5 regardless of order at which the false positive was detected. It should be noted that these false positives do create cuts in the data, similar to that of a planet and therefore affect the detection order. However, without reordering these systems we artificially inflate our multiplicity calculation in Section \ref{sec:like}. Higher multiplicities are especially sensitive to mild increases as their detection probabilities are very low. Further discussion in Section \ref{sec:inject} shows that we use the same detection efficiency for all planets found after the first detected planet, thus only planets artificially being re-assigned to 1 are of concern. Since most false positives provide relatively weak signals, only 14 systems experience this artificial re-ordering. After making the discussed cuts we find that the highest detection order existing in the parameter space is 7. This means that the highest system multiplicity we consider in this study is a 7 planet system. We find 3062 KOIs meet the indicated period and radius requirements.\

It has been suggested that gas giants eject companion planets while migrating inward \citep{bea12}. Their large Hill radius forces the planets to become unstable as the Hill radius ratio falls below $10$. These hot Jupiters create an independent population of single planet systems \citep{stef12}. If it forms via a distinct channel, this population has the ability to skew the inferred distribution of the model for the generic underlying population. To minimize such contamination, we remove all single planet systems with $r>6.7 r_{\earth}$ as indicated by \citet{stef12}. Further evidence of this independent population was discussed by \citet{joh12}, who showed that multi-planet systems with one planet of mass $>0.1$ Jupiter mass are dynamically unstable on short timescales. This $0.1$ Jupiter mass limit roughly corresponds to the $r=6.7r_{\earth}$ limit used here. We find that 120 of these single hot Jupiters exist in the dataset, leaving us with 2942 KOIs that fit all the parameter requirements described. Our final catalog of planets and their corresponding parameters can be found online.\footnote{\url{https://github.com/jonzink/ExoMult} \label{git}}\

\setcounter{table}{0} 
\begin {table*} 
\caption {The $\Gamma$ function parameters used to fit the recovery CDF displayed in Figure \ref{fig:mes}.\label{tab:mes}} 
\begin{tabular*}
	{1 \textwidth}{@{\extracolsep{\fill}}l c l c l c l c l c l} \hline \hline \multicolumn{1}{c}{Period Range} & \multicolumn{1}{c}{Maximum Detection ($c$)} & \multicolumn{1}{c}{Shape ($a$)} & \multicolumn{1}{c}{Scale ($b$)} & \multicolumn{1}{c}{Offset ($x_0$)}\\
	\hline {\bf $\Gamma_{CDF}^{m=1}$} \\
	$.5<p<200$ days & $0.9825$ & $29.3363$ & $0.2856$ & $0.0102$ \\
	$200<p<500$ days & $0.9051$ & $18.4119$ & $0.3959$ & $1.0984$ \\
	\\
	{\bf $\Gamma_{CDF}^{m \geq 2}$} \\
	$.5<p<200$ days & $0.9276$ & $21.3265$ & $0.4203$ & $0.0093$ \\
	$200<p<500$ days & $0.7456$ & $5.5213$ & $1.2307$ & $2.9774$ \\
	
	\hline 
\end{tabular*}
\end{table*}

\section{Injection Recovery}
\label{sec:inject}
Here we shall discuss how we can account for the detection efficiency as a function of detection order. \citet{chr17} injected artificial planet signals into the calibrated pixels of each of the \emph{Kepler} field stars and processed the altered light curves with the standard detection pipeline. This allows the recovery fraction to be assessed, producing a probability function based on transit MES (Multiple Event Statistic; a detailed description of MES can be found in equation \ref{equ:mes}). A $\Gamma_{CDF}$ (Cumulative Distribution Function) was fit to the empirical probability of recovery, of the form: 

\begin{equation}
	\Gamma_{CDF}(MES)=\frac{c}{b^{a}(a-1)!}\int_{0}^{MES} (x-x_0)^{a-1}e^{\frac{-(x-x_0)}{b}} dx 
\end{equation}

The purpose of this test was to establish an average detection efficiency function for the \emph{Kepler} pipeline as determined by the properties of the target star sample. Therefore planet detection order was not considered. However, many of the target stars are known to host real KOIs, and these signals will remain in the \citet{chr17} analysis. This provides an opportunity to consider the effects of detection order on recovery. Here we define detection order by the variable $m$, where $m$=1 indicates the first planet discovered in the system (i.e. highest MES). Likewise, planet $m$=2 and $m$=3 corresponds to the second and third planets found by the \emph{Kepler} pipeline. The highest detection order existing in the parameter space is 7, thus we shall work in the range of $m=1:7$.\ 

We split the data from \citet{chr17} into injection with a $.5<p<200$ days or $200<p<500$ days. The break at 200 days was selected by testing different values. Beyond 200 days, we find that the distributions begin to change significantly. To focus on the relevant parameter space of our study, we remove all injections with periods beyond 500 days and only consider stars within $4200K <T_{\rm eff}<6100K$ and $log(g)\ge4$.\

Because the goal for the original \citet{chr17} experiment was to find an overall detection probability, only one artificial signal was injected into each light curve with a radius and period uniformly sampled from $.25-7r_{\earth}$ and $.25-500$ days respectively. To understand the effect of multiple planets we therefore need to investigate systems with existing transit signals in the light curve. Over 30,000 unique signals are found within the \emph{Kepler} data pipeline. Although most of these were later deemed false positives by external checks, the pipeline treats them no differently than an actual planet. It is even very likely that some of them are in fact ``real'' planets. Therefore, injections in these systems will be subject to the same systematic issues as that of an actual multiple-planet system. This offers a far greater number of $m\geq2$ injections than those provided by the KOI list alone. For the system with injected signals, for $200<p<500$ days we find 2,099 $m\geq2$ systems and 1,579 $m\geq2$ systems for $.5<p<200$ days. We also separated the injections into $m\geq3$, but this data is extremely limited and cannot produce meaningful results without further injections. Thus, we shall focus only on m$=1$ and $m\geq2$ systems. The data are then binned in MES and the recovery fraction is determined at each binned region of MES space. Because the available data is relatively small compared to the original number of primary injections (31,302 for $200<p<500$ days and 29,083 for $.5<p<200$ days), a smoothing technique is utilized. The bin width is set to 2 MES, but instead of moving each bin by steps of width 2, the bins were recalculated at steps of 0.01 MES. This produced 800 data points across a parameter space of 0-16 MES. Utilizing this technique avoids artifacts produced when binning smaller data samples. One issue that can arise from such smoothing is an artificial distribution skew. In acknowledging this possibility, we have tested various bin widths while smoothing and find little deviation from the results with the adopted binning. Since each injection within a bin can have two possible outcome, a detection or a failed detection, the distribution within each bin will follow a binomial model. Here the number of trials corresponds to the number of injections within the bin. Thus, the uncertainty for each bin is calculated assuming a binomial distribution. The recovery CDF is then fit with a 4-parameter $\Gamma$ distribution using a $\chi^2$ minimization. The results of the fit can be seen in Table \ref{tab:mes} and Figure \ref{fig:mes}.\ 

One of the main motivations for creating these additional detection efficiency curves was a preliminary search of the results of the \citet{chr17} injection test. This showed that 61 previously detected KOIs were lost when the injection of additional planets was made. Thirty-nine of these KOI planets had a low ``Disposition Score'', indicating that small perturbation to the light curve could easily disrupt their detectability. One KOI was lost because of transit interference, where a higher MES injection with overlapping transits caused some of the transits of the weaker signal planet to be missed. Twenty-one systems had indicated that the harmonic fitting function was triggered when the injection was made, likely overfitting to the transits themselves. Ten of these light curves had no detections of planets at all. Both the injection and the KOI were missed when the artificial planet was placed into the system. This indicates that multiple-planet systems are constrained by additional detection biases not experienced by single planet systems.\    

\section{Effects of mutual inclination}
\label{sec:incl}
Here, we shall discuss how the effects of mutual inclination are handled within our model. The initial recovery study \citep{chr17} was performed without consideration of higher multiplicity planets. Thus, there was no accounting for mutual inclination. The artificial planets were injected with a random impact parameter (b) from 0 to 1. To understand the effects of mutual inclination on detection efficiency we look at the difference of impact parameters ($\Delta b$) for recovered planet systems. $\Delta b$ is calculated by taking the difference of the artificial planet and the largest MES KOI impact parameter in each system. Since an existing KOI is required for this test, we only look at systems with known planets. We find that the detected planets do not significantly differ in $\Delta b$ than the difference of two randomly drawn populations of $b$ values. Because the artificial planets were injected with uniformly drawn impact parameters, we conclude that the $\Delta b$, and therefore mutual inclination, plays an insignificant role in detection efficiency. However, larger mutual inclinations can cause certain planets to geometrically  avoid transit completely.\

\subsection{Transit Probability}
\label{sec:transp}
Analytic models of transit probability have been found for double transit systems as a function of mutual inclination \citep{rag10}. However, larger multiplicity systems are more difficult and require semi-analytic models \citep{bra16}. In order to simplify our calculation, we simulate various semi-major axis to stellar radius ratios $(a_p/R_{\star})$ and look at $10^6$ lines of sight to predict the probability of transit. To determine the period population we need a function for $m$ transit probability at some semi-major axis value ($a_p$). In order to create a function for probability of transit in addition to $m-1$ other transits, it is essential that we know the distributions of exoplanet periods. Clearly, this argument is circular in nature. We deal with this issue by using a non-uniform method of sampling from the empirical period population. This is performed for detection order $m=2:7$, since the analytic probability $(R_{\star}/a_p)$ is sufficient for m=1.\

\begin{figure}
\begin{center}
\includegraphics[width=8.5cm]{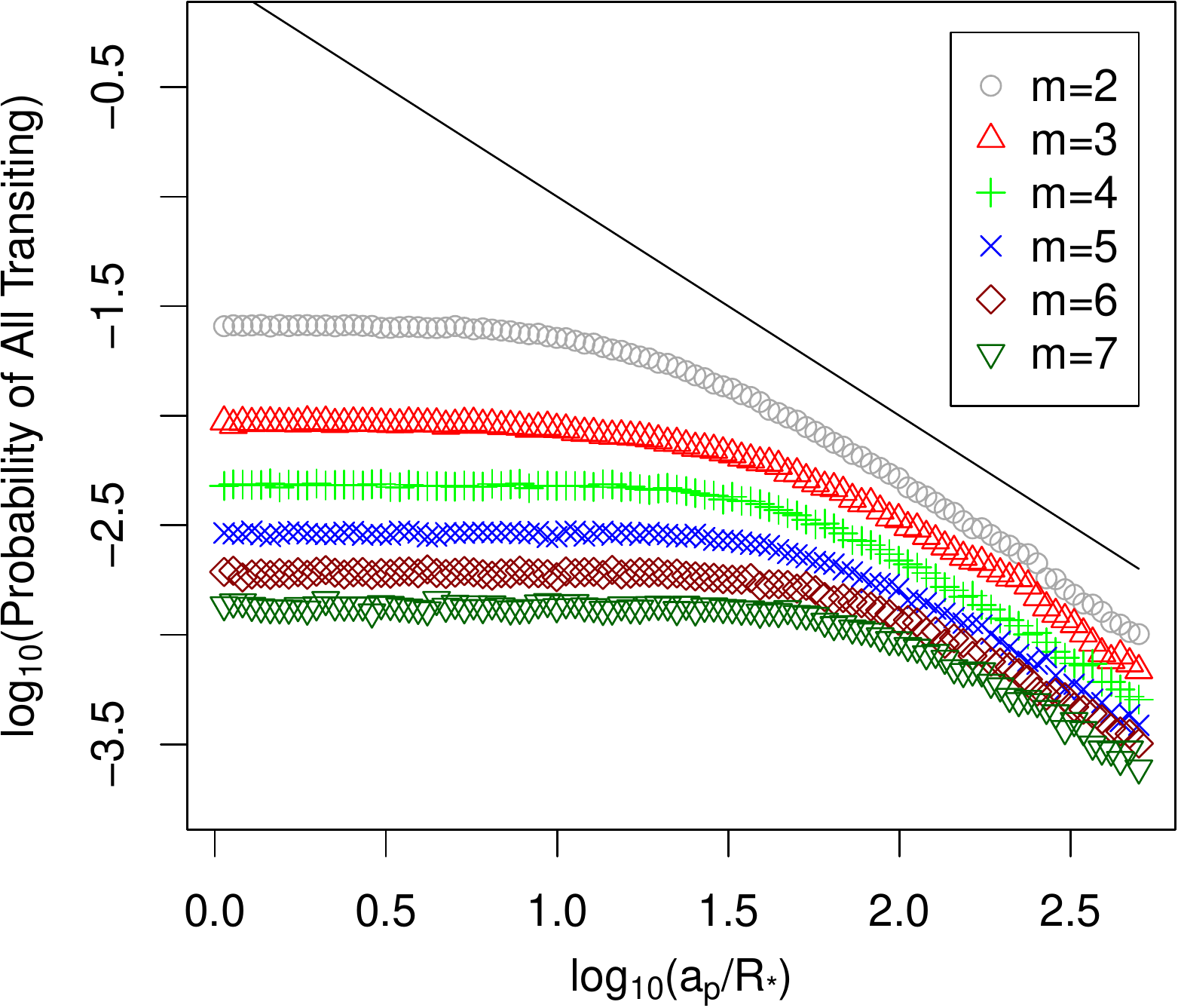}
\end{center}
\caption{The probability for transit of high multiplicity systems using the \citet{fan12} mutual inclination model. The solid black line represents the probability function used for an $m=1$ planet transit $(R_{\star}/a_p)$. A machine-readable version of this data is available \href{https://github.com/jonzink/ExoMult}{online}. \label{fig:tran}}
\end{figure}

To establish the desired detection order, the required number of planets are drawn from the empirical \emph{Kepler} period data. For example, when looking at the case of m=3, $(a_p/R_{\star})$ is selected and then the two additional planets are drawn from the known \emph{Kepler} period sample. The periods of the additional two planets are redrawn at each line of sight. This is the same as saying we marginalized the additional two planets over the \emph{Kepler} period population. In order to properly account for the transit probability of higher detection orders, we need to know the unbiased underlying populations of periods. To approximate this, we sample the empirical distribution of \emph{Kepler} planet periods, but weighted with a probability $\propto p^{2/3}$. This is done to account for the geometric bias against the detection of longer period planets. To account for the mutual inclination between orbits, we follow the $\sigma_{\sigma}$ distribution provided by \citet{fan12}. This mild distribution $(\langle\sigma\rangle=1.6^o)$ was found by looking at the impact parameter ratios within \emph{Kepler} systems. Once all orbits have been selected, the number of lines of sight where all planets transit is divided by $10^6$ to establish the transit probability. To determine whether a planet is transiting this equation must be satisfied:

\begin{equation}
\begin{split}
cos(i)*cos(\omega)-sin(i)*sin(\omega)\ge R_{\star}/a_p\\
\textup{or \space\space\space} sin(i)*sin(\omega)-cos(i)*cos(\omega)\le R_{\star}/a_p \label{equ:tran}
\end{split}
\end{equation}

where $i$ is the inclination of the of the system and $\omega$ is the ascending node. Each line of sight is drawn uniformly over $sin(i)$ and the nodes of each orbit are also drawn uniformly over $sin(\omega)$. For nodes between planets within the same system, we sample uniformly over $sin(\Delta\omega)$. We note that Equation \ref{equ:tran} is only valid for circular orbits. Consideration of eccentric orbits is presented in Section \ref{sec:ecc}. To avoid the creation of unstable systems, we check the planet separations ($|a_{p2}-a_{p1}|$). If any separation is $<10\%$ the semi-major axis of the outer planet we resample the entire system. This process is repeated until no separations fall below the $10\%$ threshold. Although mutual Hill radius would provide a better measure of stability, our metric requires no assumptions about the mass of the planets. Furthermore, we find that changing (or removing) this threshold makes little difference to the probabilities calculated, indicating that stability accounting has little effect statistically. The results of this simulation can be seen in Figure \ref{fig:tran}.\

It is worth noting that equation \ref{equ:tran} does not account for grazing transits. To properly account for this, $R_{\star}$ must becomes $R_{\star}\pm r$, where $r$ is the radius of the transiting planet. Using a uniform distribution of $r$ values from $0.5r_{\earth}$ to $16r_{\earth}$, we find that grazing transits provide an increase of 0.2\% to the overall transit probability. However, this uniform distribution is weighted far more heavily towards large planets than the underlying planet radius distribution, thus we expect the true correction to be much smaller. To properly account for grazing transit one must have some understanding of the underlying radius population. Any attempt to do so here would add more uncertainty to the calculation and provide a very minimal correction. Thus, we ignore such complications here.

\section{Detection Efficiency Grid}
\label{sec:detc}
To represent the \emph{Kepler} survey detection efficiency a grid is created in period and radius space. Both $log_{10}p$ and $log_{10}r$ are divided into 100 bins, creating 10,000 regions of the parameter space. For every region we uniformly sampled in log space for period and radius, all 86,605 stars are assigned $m$ planets based on the detection order of interest. For example, in the detection grid for the first transiting planet (m=1), the probability of detecting at least one planet is calculated at each bin. Similarly for m=2, the probability of detecting at least one planet at each bin in addition to finding another planet in some other arbitrary bin. The average detection probability for each region is calculated using these planetary assignments and the procedures provided in the next Sections (\ref{sec:probd1}; \ref{sec:probd2}). This process is then repeated for each of the 10,000 regions. We calculated 7 detection efficiency grids: first planet probability (m=1), second planet probability (m=2),..., and the seventh planet probability (m=7). This procedure is similar to that of \citet{bur15} and \citet{tru16}, but now with 7 different detection order grids. \

\begin{figure*}
\centering
\hfill \raisebox{0.21\height}{\includegraphics[width=.33cm]{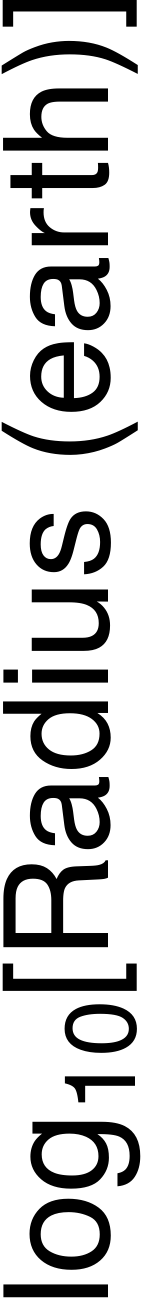}} \hfill \includegraphics[width=3.90cm]{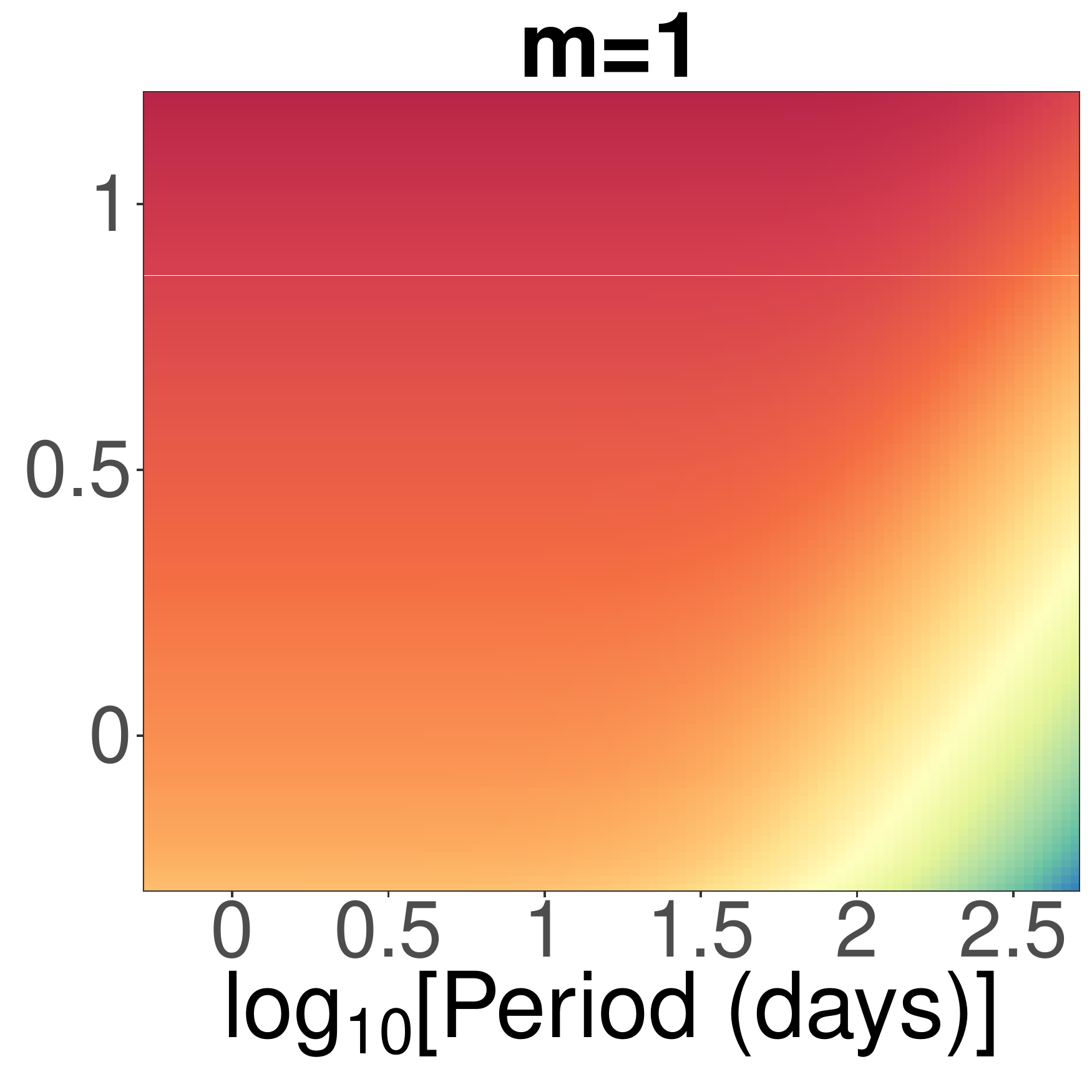} \hfill \includegraphics[width=3.90cm]{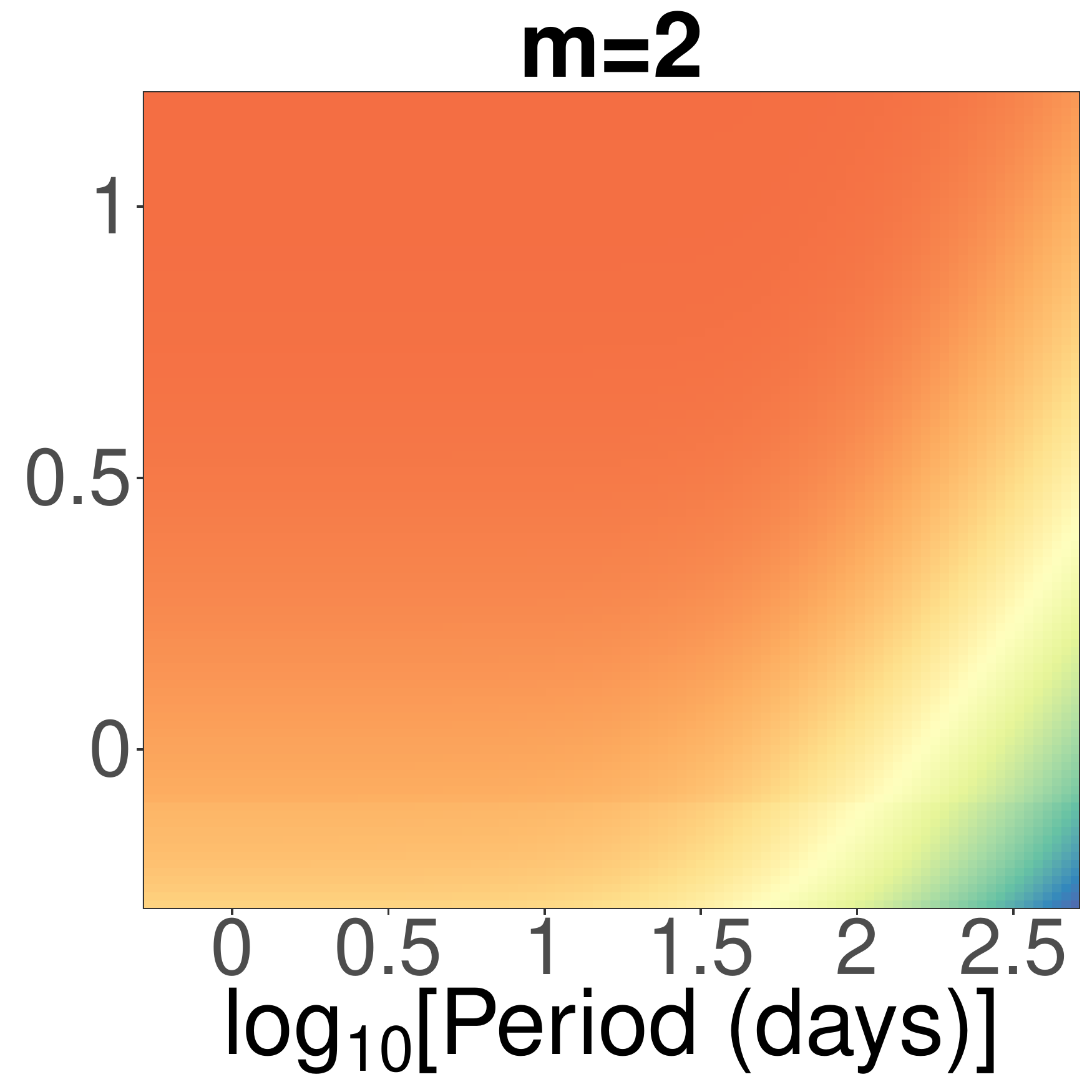} \hfill \includegraphics[width=3.90cm]{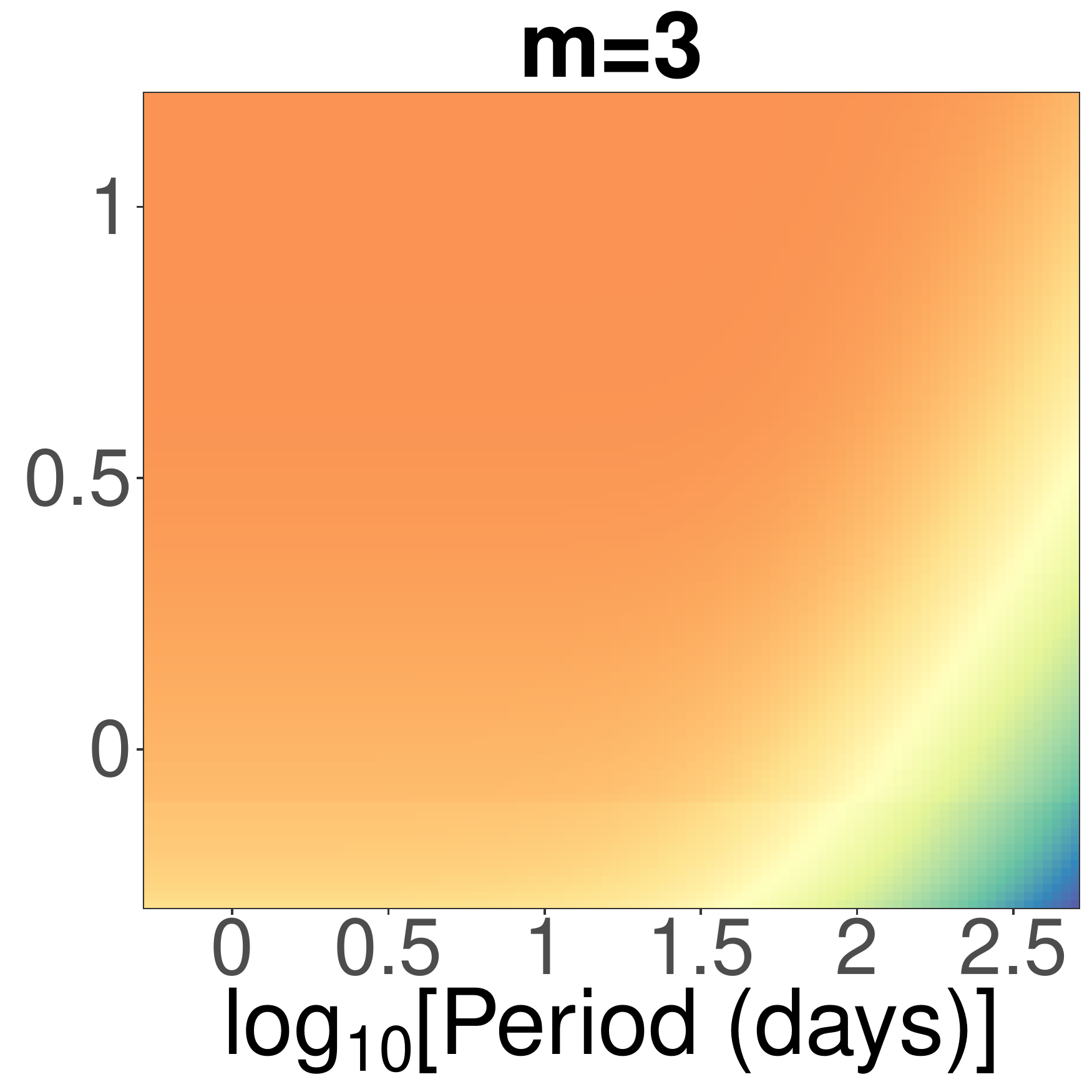} \hfill \includegraphics[width=3.90cm]{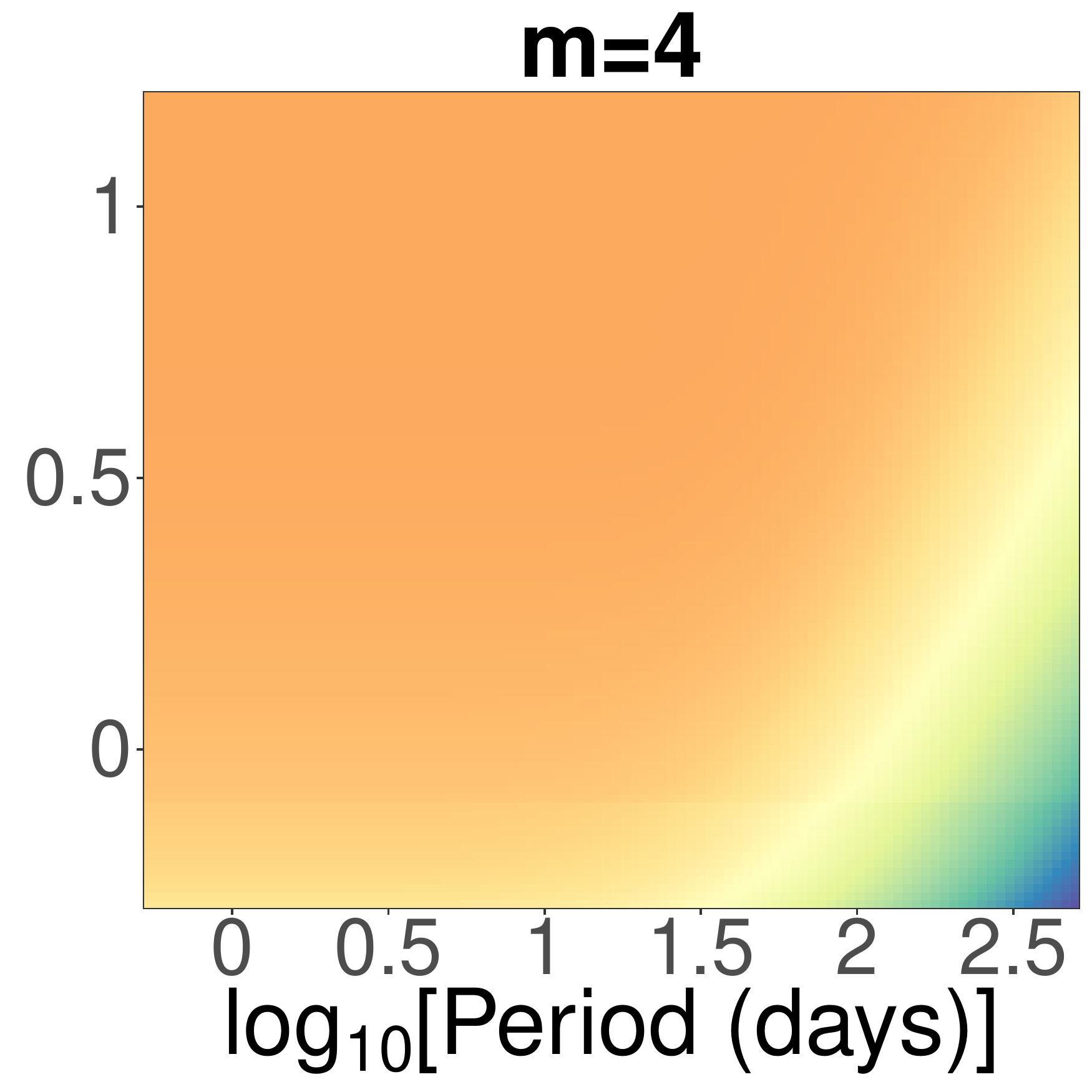} \hfill \includegraphics[width=1.20cm]{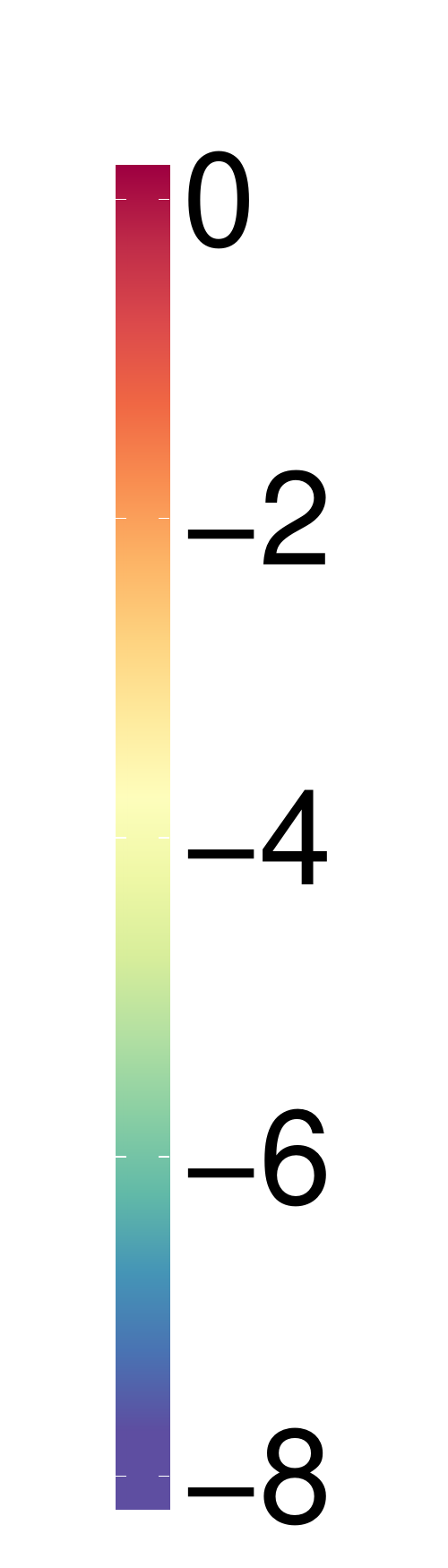} \hfill \caption{The detection efficiency maps for m=1:4 exoplanet discovery orders. The color map is representative of $log_{10}($Detection Probability$)$. The fading of color across detection order (m) shows the decreasing detection probability. A machine-readable version of this data is available \href{https://github.com/jonzink/ExoMult}{online}. \label{fig:heat}}
\end{figure*}

\subsection{Probability of Detection for $m=1$}
\label{sec:probd1}
We shall begin with the formula for the detection of the first planet and then discuss the modifications made for the detection of higher order systems. In our base model we assume all planets have perfectly circular orbits and consider the effects of eccentricity in Section \ref{sec:ecc}. This assumption of little or no eccentricity is reasonable for the typical multiple systems sampled by \emph{Kepler}, where non-circular orbits would result in unstable system architecture. To account for the geometric probability of transit we use: 

\begin{equation}
P_{tr}=\frac{R_{\star}}{a_p} 
\end{equation}

where $R_{\star}$ is the radius of the star and $a_p$ is the semi-major axis of the planet orbit. The chord at which the planet transits across the stellar host is given by 

\begin{equation}
f_{tr}=\sqrt{1-b^{2}} 
\end{equation}

where $b$ is the impact parameter of the planet transit. $b$ is assigned by uniformly sampling between 0 and 1 for each planet. The duration of the transit can be calculated as
 
\begin{equation}
t_{dur}=\frac{R_{\star}*f_{tr}}{a_p*\pi} (\frac{p}{1 \textup{day}})*24 \textup{hr} 
\end{equation}

where $p$ is the orbital period of the planet. The expected number of transits can be found with 

\begin{equation}
n_{tr}=\frac{data_{span}}{p} 
\end{equation}

where $data_{span}$ is the span of the data within the \emph{Kepler} survey. Because of various shut downs and data downloads throughout the \emph{Kepler} mission, it is possible that some of the transits may have been missed. To account for the probability of the transit occurring in the window of the \emph{Kepler} mission we adopt the window function provided by \citet{bur15}.

\begin{equation}
j=\frac{data_{span}}{p}
\end{equation}

\begin{equation}
\begin{split}
P_{win}=1-(1-duty)^{j}-j*duty(1-duty)^{j-1}\\
-\frac{j(j-1)duty^{2}(1-duty)^{j-2}}{2} \label{equ:pwin}
\end{split}
\end{equation}

where $duty$ is the duty fraction of the targeted stellar source. The \emph{Kepler} pipeline requires at minimum 3 transits for candidate consideration; $P_{win}$ is the probability that at least 3 transits will be detected by the available \emph{Kepler} data. Since most targets have a $duty=.95$, short period transits ($j>>3$) produce a $P_{win}$ nearly 1 and approach 0 as $j<3$. Almost all of our sample have data  throughout the full data set span of 1458.931 days. The mean $data_{span}$ for this study is 1427.445 days.\

Other studies have used various way to account for the effects of limb darkening such as that of \citet{cla11}.  We attempt to mimic the pipeline by looking at the empirical limb darkening values chosen for existing KOIs (with the same stellar parameters discussed in Section \ref{sec:stell}). We find that the two limb darkening parameters ($u_1,u_2$) used to fit planet transits within the pipeline are strongly correlated to stellar temperature ($T_{\rm eff}$). The best fit line to this correlation is as follows: 

\begin{equation}
\begin{split}
u_1&=-1.93*10^{-4}*T_{\rm eff}+1.5169 \\
u_2&=1.25*10^{-4}*T_{\rm eff}-0.4601 
\end{split}
\end{equation}

We warn that these correlations mimic the choice of the pipeline rather than the true stellar features and should not be used for more evolved stars with $log(g)<4$. With the given calculated parameters, it is now possible to calculated the expected MES of the \emph{Kepler} pipeline as presented by \citet{bur17a}. 

\begin{equation}
k_{rp}=\frac{r}{R_{\star}}
\end{equation}

\begin{equation}
c_0=1-(u_1+u_2)
\end{equation}

\begin{equation}
\omega=\frac{c_0}{4}+\frac{u_1+2*u_2}{6}-\frac{u_2}{8}
\end{equation}

\begin{equation}
\begin{split}
depth=1-(\frac{c_0}{4}+\frac{(u_1+2*u_2)*(1-k_{rp}^2)^{\frac{3}{2}}}{6}\\
-\frac{u_2(1-k_{rp}^2)}{8})\omega^{-1}
\end{split}
\end{equation}

\begin{equation}
MES=\frac{depth}{CDPP*10^6}*1.003*n_{tr}^{\frac{1}{2}} \label{equ:mes} 
\end{equation}

where $CDPP$ is in ppm from the \emph{Kepler} stellar catalog, interpolated by the transit duration.  Finally, we account for the systematic detection efficiency using the Gamma distribution CDF described in Section \ref{sec:inject}. 

\begin{equation}
P_{tip}^{m=1}=\Gamma_{CDF}^{m=1}(MES) 
\end{equation}

where the parameters for $\Gamma_{CDF}$ are the given in Table \ref{tab:mes}. Combining all of the discussed probabilities provides an estimate of the detection likelihood of the highest MES planet within the system. This probability is given as follows: 

\begin{equation}
P_{det}^{m=1}=P_{tr}*P_{win}*P_{tip}^{m=1} 
\end{equation}

This equation provides a metric for understanding the bias of the highest MES planet. This probability is dependent on detection order and we shall now discuss in the next section how higher multiplicity planets ($m\ge2$) can be accounted for.

\subsection{Probability of Detection for $m\ge2$}
\label{sec:probd2}
For $m\ge2$ planets we follow much of what is described in the previous Section (\ref{sec:probd1}), with a few mild changes to better model the differences in detection probability.\

We change the transit probability to reflect the probability of $m$ planets transiting, accounting for the probability of finding this planet with at minimum $m-1$ other planets. To best capture the probabilities of our simulation in Section \ref{sec:transp}, we interpolate between simulated data points for the transit probability. 

\begin{equation}
P_{tr}^m=\textup{Linear Interpolate}(m,\frac{a_p}{R_{\star}}) 
\end{equation}

For example, if we are looking at a planet with m=3 (the third planet detected) with $a_p/R_{\star}=32$, we would expect a transit probability of $\sim0.008$. This can be clearly seen in the data provided by Figure \ref{fig:tran}. Since no such simulated value exist at this exact point, we interpolate between the the two neighboring estimations to establish this value. Here we use the new detection efficiency for higher m planets. 

\begin{equation}
P_{tip}^{m\ge2}=\Gamma_{CDF}^{m\ge2}(MES) 
\end{equation}

\begin{equation}
P_{det}^{m}=P_{tr}^m*P_{win}*P_{tip}^{m\ge2} 
\end{equation}

where equation \ref{equ:pwin} is again used for $P_{win}$. In reality, there are differing window functions for each detection order; when tested, we find that $\approx 0.4\%$ of the light curve is lost with the addition of each planet. One can see that varying the $duty$ parameter of equation \ref{equ:pwin} by even $3\%$ has negligible effects on the $P_{win}$ value. Because the detection efficiency is the same for $m\ge2$, the only difference between the $m=2:7$ probability maps is the transit probability. This now produced 7 distinct detection grids ($m=1:7$). The first four grids can be seen in Figure \ref{fig:heat}. The detection order of the exoplanet in question will dictate which grid is most appropriate for application. \  

To summarize, we have described how the recovery probabilities (CDF) are a function of detection order (m). We use this to create 7 different detection efficiency maps (Figure \ref{fig:heat}). In order to create a map for m=1 planets, we sample across planet period and radius space. Doing so, we calculate the probability of detection and averaged over all stars within the \emph{Kepler} stellar sample. We expand upon this idea, creating a map for m=2 planets. Here the new recovery CDF is implemented to account for the additional loss of planets at higher detection orders. Furthermore, we account for the probability of two planets within the system transiting using a mild mutual inclination model (Figure \ref{fig:tran}). Jumping from m=1 to m=2 we lose an additional 5.5\% and 15.9\% of the planets for periods $<200$ days and periods $>200$ days respectively. This is due to properties of the pipeline when fitting multiple transit systems. This procedure is repeated for m=3:7 each accounting for the appropriate number of transiting planets according to the data in Figure \ref{fig:tran} (3-7 respectively). There is an additional loss of nearly 70\% at each respective discovery order due to the unlikely event of multiple orbital alignment with our line of sight. It is clear that these two factors, geometric transit likelihood and pipeline recovery, have a significant effect on the multiplicity extracted from the \emph{Kepler} data set.\

\begin{figure}
\begin{center}
\includegraphics[width=8.5cm]{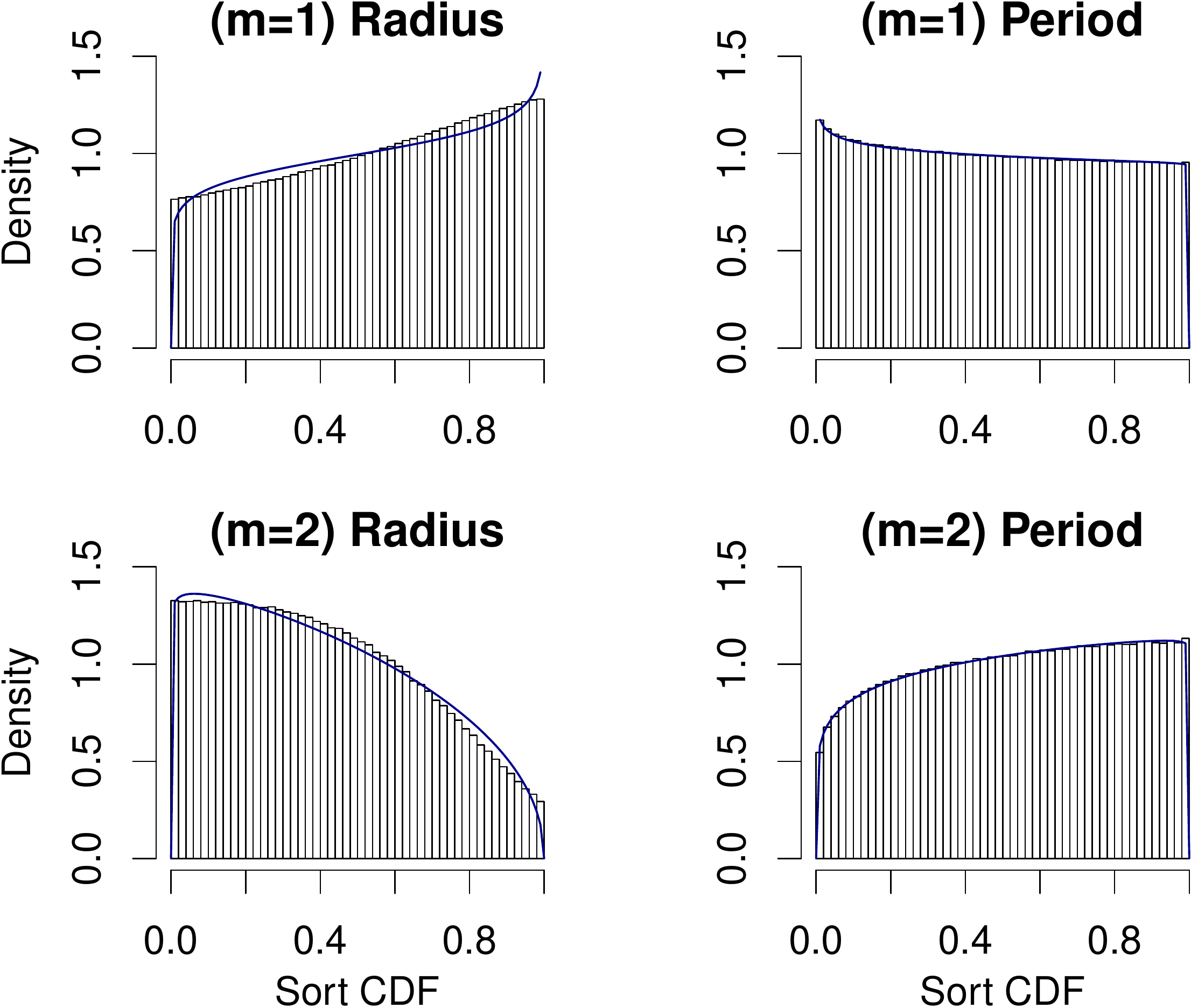}
\end{center}
\caption{The sorting simulation for m=1 and m=2. The solid blue line represents the Beta distribution fit to the respective data set. The boxes are a histogram of the simulated data after being sorted. It is apparent that sorting has a more dramatic effect on radius than period. This is expected as $MES \propto r^2/p^{1/3}$. A mild deviations from the model is noted in the radius skew. This discrepancy dissolves as we move into higher detection orders. Furthermore, the effects of these deviations are insignificant, given the cuts on duty cycle, data span, and stellar type already made. \label{fig:sort}}
\end{figure}

\begin {table*} 
\caption {The parameters found in testing the sorting effects of MES. These parameters correspond to a Beta distribution skew expected for the CDF of each multiplicity population. \label{tab:sort}} 
\begin{tabular*}
{1 
\textwidth}{@{\extracolsep{\fill}}l c l c l c l c l c l c l c l c l} \hline \hline \multicolumn{1}{l}{} & \multicolumn{1}{l}{m=1} & \multicolumn{1}{l}{m=2} & \multicolumn{1}{l}{m=3} & \multicolumn{1}{l}{m=4} & \multicolumn{1}{l}{m=5} & \multicolumn{1}{l}{m=6} & \multicolumn{1}{l}{m=7}\\
\hline
$a_{rad}$ & 1.095 & 1.030 & 1.028 & 1.013 & 0.998 & 1.065 & 0.951\\
$b_{rad}$ & 0.923 & 1.470 & 2.206 & 3.063 & 4.013 & 4.898 & 6.614\\
$a_{per}$ & 0.957 & 1.152 & 1.172 & 1.184 & 1.183 & 1.166 & 1.234\\
$b_{per}$ & 1.004 & 1.010 & 1.000 & 0.999 & 0.997 & 1.006 & 0.994\\
\hline 
\end{tabular*}
\end{table*}

\section{The Likelihood Function}
\label{sec:like}
Using the efficiency grids derived in the previous section, we can infer properties of the underlying planetary population. Here we will discuss the likelihood function required to implement Bayes theorem and extract these population parameters.\

We adopt the approach of previous studies (e.g. \citealt{you11,pet13a,bur15}), modeling the underlying population as characterized by independent power-law distributions
in period and radius. We also make explicit the assumption that there is a single planetary population -- assuming that systems which show only one transit are drawn
from the same underlying distribution as those which show multiple transits. We will examine the validity of this assumption in Section \ref{sec:mult}. Our focus on multiple
systems also means that we include more of the \emph{Kepler} parameter space than was used in most previous papers.

The population of exoplanets is modeled as follows:

\begin{equation}
\frac{d^2N}{dpdr}=fg(p)q(r) 
\end{equation}

\begin{equation}
g(p)= 
\begin{cases}
C_{p1}p^{\beta_1} & \text{if $p<p_{br}$} \\
C_{p2}p^{\beta_2} & \text{if $p \geq p_{br}$} \\
\end{cases}
\end{equation}

\begin{equation}
q(r)= 
\begin{cases}
C_{r1}r^{\alpha_1} & \text{if $r<r_{br}$} \\
C_{r2}r^{\alpha_2} & \text{if $r \geq r_{br}$} \\
\end{cases}
\end{equation}

where $f,\alpha_1,\alpha_2,\beta_1,\beta_2,p_{br},$ and $r_{br}$ are all fit parameters. We require continuity at $r_{br}$ and $p_{br}$ through the normalization constants for $q(r)$ and $g(p)$.\

Our method expands on the Poisson process likelihood used by \citet{you11}. The main difference is the separation of planets by detection order ($m$). In doing so, we require different occurrence factors ($f$) for each $m$, increasing the required number of parameters. Previous studies such as \citet{bur15} have used a single occurrence value, providing an average occurrence factor. By separating the occurrence factor as a function of detection order, we can allow for differences in detection efficiency while simultaneously fitting for the occurrence of planet multiplicity. 

\begin{equation}
Likelihood = \prod_{m=1}^{7} [\prod_{i=1}^{n_{m}}f_m\eta_{m}(p_i,r_i) g(p_i)q(r_i)]e^{-N_{m}} 
\end{equation}

\begin{equation}
\begin{split}
&N_{m}=\\
&86,605 f_m \int_{.5 \textup{days}}^{500 \textup{days}}\int_{.5 r_{\earth}}^{16 r_{\earth}} \eta_{m}(p,r) O_m(p_i,r_i) g(p) q(r) dr dp
\end{split}
\end{equation}

where $N_m$ represents the expected number of planets detected for each discovery order ($m$) and $f_m$ is an occurrence factor for each $m$. This value provides information on the occurrence of each $m$ multiplicity. However, to find meaningful information from these values, they must be disentangled from each other as discussed in Section \ref{sec:occ}. The 86,605 accounts for the number of stars in our test sample and $\eta_{m}(p,r)$ is the detection probability at the given detection order. The function $O_m(p_i,r_i)$ is the sorting order correction for the (PDF) Probability Distribution Function. This function is necessary to account for the bias in the PDF introduced by our sorting in terms of detection order (discussed further in \ref{sec:sort}).\

It is often more useful to consider the natural log of the likelihood, which can be simplified: 

\begin{equation}
ln (Likelihood) \propto \sum_{m=1}^{7} [\sum_{i=1}^{n_{m}}ln(f_m g(p_i)q(r_i))]-N_{m} 
\end{equation}

Using the $ln(Likelihood)$ is common practice with fitting algorithms, where the ratio of likelihoods are compared to determine the best fit (maximum likelihood). Since $\eta_{m}(p,r)$ is not dependent on the fitting parameters it can be considered constant.

\subsubsection{Calculating $N_{m}$}
To find $\eta_{m}(p,r)$ we use the detection maps found in Section \ref{sec:probd1} and \ref{sec:probd2}. Here we are assuming an average probability of detection over the stellar population. To properly treat this integral, one would have to compute the detection probability for each star. Such a procedure would be computationally expensive and provide a minimal increase in precision.\

\begin {table*} 
\caption {The mixture probabilities for each detection order. For example, this accounts for the possibility that two and three planet systems may only be found with a single planet. These values were found using our transit probability model described in Section \ref{sec:transp}. \label{tab:entag}} 
\begin{tabular*}
{1 \textwidth}{@{\extracolsep{\fill}}l c l c l c l c l c l c l c l} \hline \hline \multicolumn{1}{l}{} & \multicolumn{1}{l}{m=1} & \multicolumn{1}{l}{m=2} & \multicolumn{1}{l}{m=3} & \multicolumn{1}{l}{m=4} & \multicolumn{1}{l}{m=5} & \multicolumn{1}{l}{m=6}\\
\hline $ \frac{P(2|\overline{(m,1)})}{P(m)}$ & 0.67 & - & - & - & - & - \\
$\frac{P(3|\overline{(m,2)})}{P(m)}$ & 0.68 & 0.50 & - & - & - & - \\
$\frac{P(4|\overline{(m,3)})}{P(m)}$ & 0.53 & 1.05 & 0.50 & - & - & - \\
$\frac{P(5|\overline{(m,4)})}{P(m)}$ & 0.53 & 1.12 & 1.52 & 0.46 & - & - \\
$\frac{P(6|\overline{(m,5)})}{P(m)}$ & 0.37 & 1.07 & 1.85 & 1.69 & 1.22 & - \\
$\frac{P(7|\overline{(m,6)})}{P(m)}$ & 0.33 & 0.71 & 1.64 & 1.90 & 1.25 & 1.22\\

\hline 
\end{tabular*}
\end{table*}

\subsubsection{Sorting Order} 
\label{sec:sort}

Here we will provide a brief overview of order statistics and why it is an important feature of this model. As mentioned previously, the \emph{Kepler} pipeline finds planets in order of decreasing MES. Such ordering will skew the distribution of planets found in each $m$. Larger, short period, planets will tend to be found in order $m=1$ or $m=2$, because there are more transits and deeper transit depths. Smaller, long-period planets will tend towards orders $m=6$ or $m=7$. To account for such a skew, a joint distribution model ($P_m(x)$) can be utilized \citep{dav03}. 

\begin{equation}
P_m(x) \propto P_0(x)C_0(x)^{a_m-1}(1-C_0(x))^{b_m-1} 
\end{equation}

Here, $P_0(x)$ is the true underlying probability distribution function and $C_0(x)$ is the true cumulative distribution function. $a_m$ and $b_m$ can range from (0,$\inf$) and forces the skew of the distribution. Essentially, the PDF of the distribution is skewed by a Beta distribution of the CDF. In the case of $a_m=b_m=1$ the sorting skew returns the original PDF ($P_0(x)$).\ 

The parameters $a_m$ and $b_m$ can be found analytically for equally sampled orders, but becomes far more complex in the decreasing case at hand (each $m$ has fewer planets than the last). To determine the best values for this case, we choose to simulate this sorting mechanism on a uniform distribution, where the skew can be clearly isolated and extracted. In doing so, we force  the ratio of each $m$ sample to mimic that of the empirical population. Each system is then sorted by $r^2/p^{1/3}$, imitating \emph{Kepler's} MES sorting. For example, if a system of (r=1.2,p=25), (3.5,20), (4.1,150) were randomly drawn into $m=1,2,3$ detection orders, they would be re-sorted as (3.5,20), (4.1,150), (1.2,25) corresponding to $m=1,2,3$. As we can see, the highest MES will always rise to $m=1$. This is then repeated for $10^7$ systems. Figure \ref{fig:sort} shows how the first two detection orders are skewed by this procedure. If sorting were not an issue, these distributions would maintain the uniform flat appearance. Fitting a Beta distribution to this skew, we can determine the best $a_m$ and $b_m$ parameters for our sample. These parameters are provided in Table \ref{tab:sort}. \

Since the this joint distribution is separable, we define the skew portion of the distribution as $O_m(p,r)$. 

\begin{equation}
\begin{split}	
O_m(p,r) = N &* C_r(r)^{a_{m,r}-1}[1-C_r(r)]^{b_{m,r}-1}\\
 			 &* C_p(p)^{a_{m,p}-1}[1-C_p(p)]^{b_{m,p}-1} 
\end{split}
\end{equation}

where $C_r(r)$ and $C_p(p)$ represents the CDF of the radius and period distributions respectively and $N$ represents a normalization factor that we find numerically within the MCMC. 

\subsubsection{Occurrence Factor} 
\label{sec:occ}
As noted, the value $f_m$ is an integrated occurrence factor. In order to extract meaningful values, we realize that many $m=6,7$ planet systems will only provide detectable transits for one or two planets within the system. This will lead to an increased contribution to lower detection orders. Thus we adopt the following method for disentangling the true occurrence factors ($F_m$): 

\begin{equation}
f_m=F_m +\sum_{n=m+1}^{7} F_n \frac{P(n|\overline{(m:n-1)})}{P(m)} 
\end{equation}

Here $P(n|\overline{(m:n-1)})$ represents the probability of finding planet $n$ given that planets $(m:n-1)$ are not found and $P(m)$ is the probability of finding planet $m$. This ratio accounts for the dependence between occurrence factors. If the mutual inclination is purely isotropic and planets are truly independent this ratio would be one. We use our transit simulation from Section \ref{sec:transp} to extract these marginalized probabilities. Table \ref{tab:entag} contains the results from this simulation. This model indicates that each multi-planet system will have more than one opportunity to find an $f_1$ planet. The physical interpretation of the $F_m$ values is the fraction of stars that have at least $m$ planets.\ 

\subsection{Fitting the Data}
We employ EMCEE \citep{for13}, an affine-invariant ensemble sampler \citep{goo10}, to explore the parameter space of our study. To better constrain the 13 fit parameters, a Bayesian framework is implemented. Linear space uniform priors are used for all parameters. For $\alpha_1,\alpha_2,\beta_1,$ and $\beta_2$ the priors range from -30 to +30. For $r_{br}$ and $p_{br}$ the priors range from $r_{min}$ and $p_{min}$ to $r_{max}$ and $p_{max}$ of our planet sample respectively. One unique restriction for our prior is that $F_m$ must be larger than $F_{m+1}$. It is not possible to have a higher occurrence of $m+1$ than $m$ planets. To avoid truncation bias and maintain order, all $F_m$ priors range from 0 to $F_{m-1}$. In the special case of $m=1$, the prior ranges from 0 to 1. It is important to remember that $F_m$ represents the fraction of the population containing $m$ planets. Therefore, this cascading prior still allows for larger multiplicity systems to be more common than smaller multiplicity systems.\

\section{Discussion}
\label{sec:discuss}

In this section, we now apply the formalism we have developed to infer the revised occurrence rate parameters for planets orbiting GK dwarfs. This sample includes data from the final \emph{Kepler} release DR25 and updated planet radius measurements from the CKS and \emph{Gaia} DR2. Beyond these recent data improvements, we now include a corrected detection efficiency for multiple-planet systems. Given that many multiple-planet systems span much of the \emph{Kepler} Parameter space, we include planets within $.5<r<16 r_{\earth}$ and $.5<p<500$ days. In implementing two detection efficiencies, this study expands on the Poisson process likelihood function used by other authors, allowing for the treatment of planet multiplicity. This Bayesian framework is fit using an MCMC, where 20,000 steps are used to model the posterior of each parameter. The resulting posteriors are presented in Figure \ref{fig:post}. From this model we infer best fit power-law values of $\alpha_1=-1.65\pm^{0.05}_{0.06}$, $\alpha_2=-4.35\pm0.12$, $\beta_1=0.76\pm0.05$, and $\beta_2=-0.64\pm0.02$. The breaks in our best fit model occur at $p_{br}=7.08\pm^{0.32}_{0.31}$ days and $r_{br}=2.66\pm0.06 r_{\earth}$. \

One novel feature of our fitting method is the ability to extract exoplanet multiplicity. This information is provided through the $F_m$ parameters. These values indicate the probability of a system having at least $m$ planets. We find the following value best fit our model: $F_1=0.72\pm^{0.04}_{0.03}$, $F_2=0.68\pm0.03$, $F_3=0.66\pm0.03$, $F_4=0.63\pm0.03$, $F_5=0.60\pm0.04$, $F_6=0.54\pm^{0.04}_{0.05}$, and $F_7=0.39\pm^{0.07}_{0.09}$.\    
 
\begin{figure*}
\hfill \includegraphics[width=8.5cm]{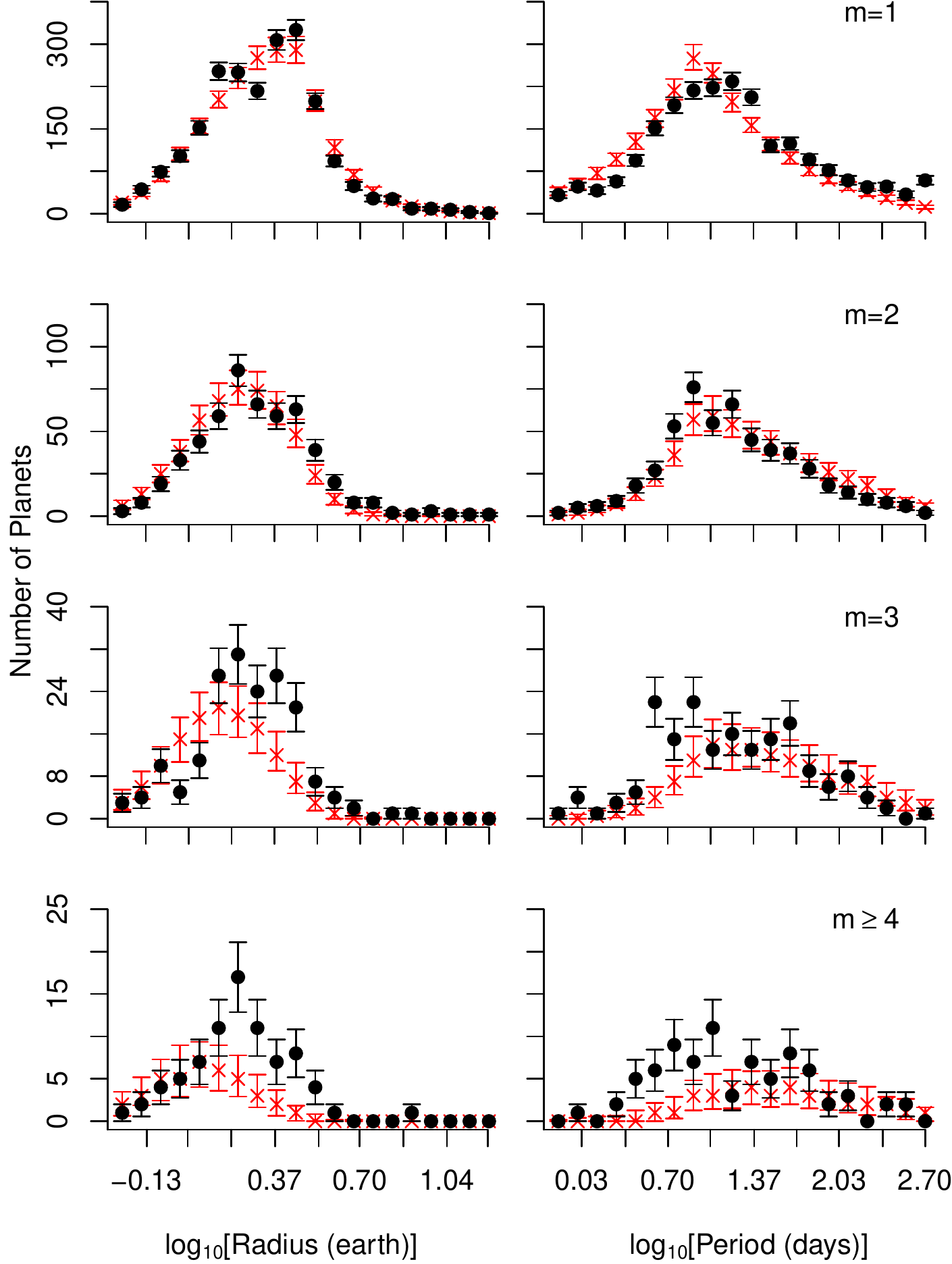} \hfill \includegraphics[width=8.5cm]{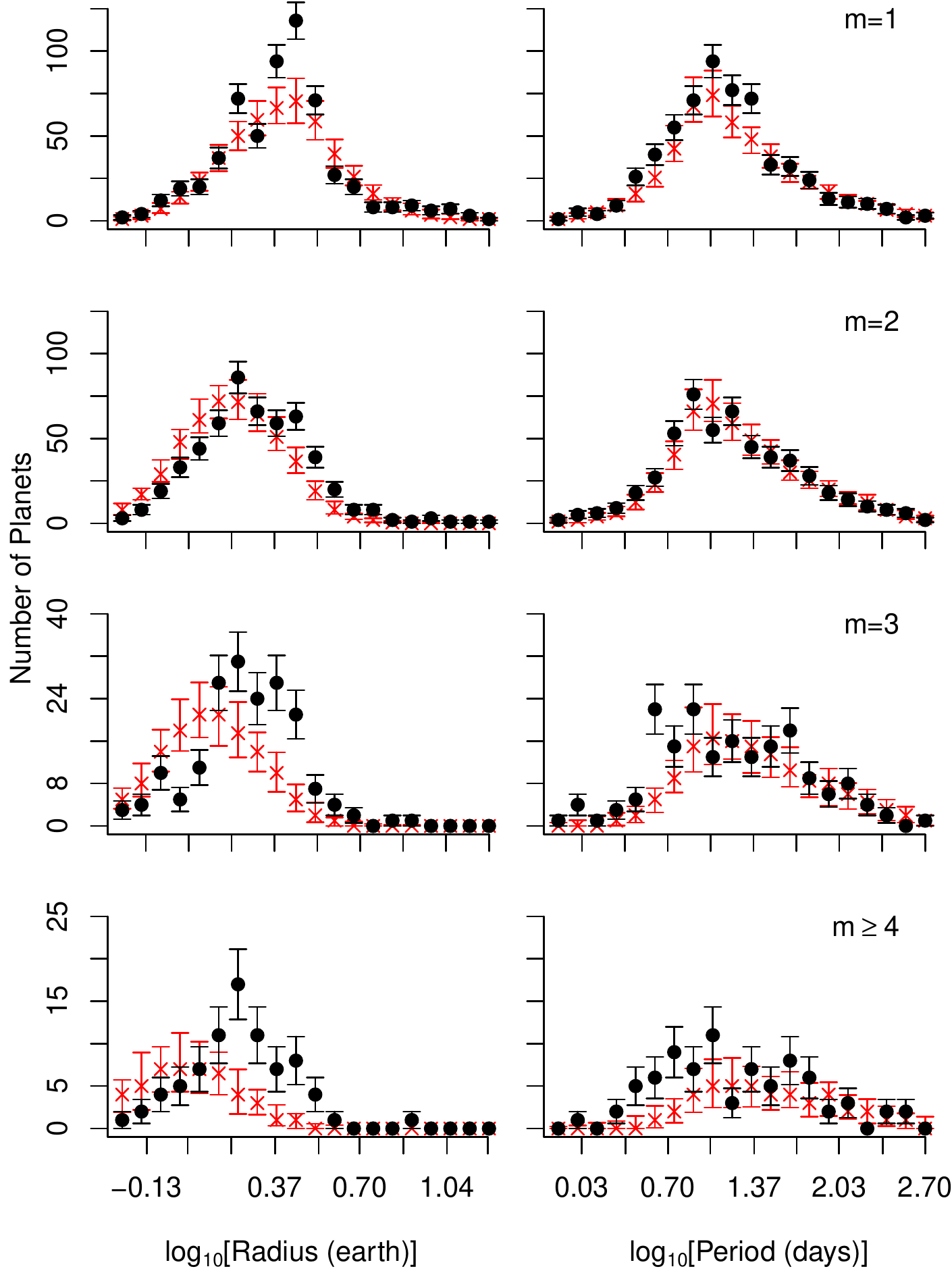} \hfill
\caption{A plot of the forward modeled population derived by our Bayesian analysis. The red x marks symbolize the model values with their corresponding 68.3\% confidence intervals. To find this interval the model is sampled 50 times using the posterior parameter distributions, the uncertainty reflects the fluctuations we find from these trials. The black points show the \emph{Kepler} data with poisson uncertainty. For m=5:7 many of the bins have 1 or 0 planets, where small number statistics cause significant variations. In order to minimize this variations we present the resulting combination of $m\ge4$. However, it should be noted that our forward model does differentiate between these detection orders. {\bf Left:} Forward model of multiple and single planet systems. {\bf Right:} Forward model of only multiple-planet systems. This model was produced by only fitting to the data of multiple-planet systems. \label{fig:for}}
\end{figure*}

\subsection{Forward Modeling the Results} 
\label{sec:mult}
Thus far, we have accounted for various parameter and population dependencies. To ensure that this process yields meaningful results, we choose to sample the extracted population and subject it to the detection constraints described in Section \ref{sec:detc}. Here we present the \textbf{\href{https://github.com/jonzink/ExoMult}{ExoMult}} forward modeling software. This code, developed in \emph{R}, simulates these detection effects and produces a population of detected planets. Using this program, we can make far fewer assumptions and directly recover the expected population. For example, the probability of transit for all 7 planets can be directly accounted for by sampling system inclination, mutual inclination and the argument of periapsis directly. Furthermore, the detection probability will not be marginalized over all stars, but rather reviewed for each system independently.\

The first step in our forward model is drawing each system of planets according to the population parameters given in Figure \ref{fig:post}. Each system is randomly oriented with mutual inclinations drawn from a Rayleigh distribution. For planets with detectable impact parameters ($b<1$), the planets within each system are sorted in decreasing MES. The probability of recovery is assigned to each planet according to the procedure laid out in Sections \ref{sec:probd1} and \ref{sec:probd2}. Based on the calculated probability of detection, the planet is either detected or lost by drawing from a random number generator. Figure \ref{fig:for} shows the best fit model to the observed population obtained with  this forward model. It is clear the our Bayesian method provides a reasonable model, where nearly all data points are within a $1\sigma$ deviation of the observed distribution.\

\citet{ful17} and \citet{ber18} have provided evidence for a dip in the radius population around $1.5-2r_{\earth}$. This gap is apparent in the m=1 case of Figure \ref{fig:for}. While the deviation from a broken power-law is mild, we explore the effects here. When we remove the single planet systems from the data set, this gap is no longer apparent. One plausible explanation for this gap is a unique population of single planet systems (Although evidence from \citet{weiss18} shows that a weak gap can be seen in the multi-planet systems when aggregated). To explore this theory, we isolate the multi-planet systems and run our fitting procedure again. We find a mild difference in the extracted $\alpha$ or $\beta$ power law values ($\alpha_1=-1.98\pm0.08$ ; $\alpha_2=-3.90\pm0.16$ ; $\beta_1=0.96\pm0.08$ ; $\beta_2=-0.79\pm0.03$). This indicates that if a separate population does exist, the population parameters are weakly affected by their inclusion in our dataset. The resulting forward model of this fit is presented in Figure \ref{fig:for}. Furthermore, the increase in uncertainty seen in these parameters is due to the reduced samples used for fitting (1305 multiple system candidates vs. 2942 total candidates). It is notable that the empirical \emph{Kepler} data set is sharply peaked, while the model does not provide a similar sharpness for the $m=1$ radius population (Figure \ref{fig:for} Right). This could be due to the existence of the mentioned radius gap. Furthermore, it is possible that a true accounting for planet period and radius covariance could produce such a peak. \citet{mil17} and \citet{wei17} show that the planets within multiple systems tend to have similar mass and radius components. Although these features are not properly accounted for here, Figure \ref{fig:for} (Left) shows that these mild population characteristics remains small and do not deviate greatly from a simple broken power-law model. We hope to include such features in the next iteration of this software.\

It is possible that future studies may use this forward modeling technique to directly determine the population parameters. Unfortunately, it remains computationally expensive to properly account for all detection features. \citet{tru16} overcame this cost by ignoring multiplicity.\

\subsection{Comparison with Prior Work} 
\label{sec:disc}
We use a Bayesian method to infer population parameters for the \emph{Kepler} exoplanet population, following much of the procedure presented in \citet{you11}. However, we build upon this method to extract information about the population multiplicity. Using a broken power-law distribution we find that population parameters of $\alpha_1=-1.65\pm^{0.05}_{0.06}$, $\alpha_2=-4.35\pm0.12$, $\beta_1=0.76\pm0.05$, and $\beta_2=-0.64\pm0.02$ provide the best replication of the empirical population. The best fit breaks in these distributions are as follows: $p_{br}=7.08\pm^{0.32}_{0.31}$ days and $r_{br}=2.66\pm0.06 r_{\earth}$.\       

Many prior studies have examined the occurrence of planets as determined by \emph{Kepler}. \citet{you11} provided an early estimate of the occurrence rate using a Poisson process likelihood, finding that the PDF exhibited a power law break at periods $\sim 7$~days, with $\alpha=-2.44$ and $\beta=3.23$ at short periods, and $\alpha = -2.93$ and $\beta=-0.37$ at longer periods (we have converted his numbers into the definitions of $\alpha$ and $\beta$ adopted here). These suggest a steep rise towards smaller radius planets at all periods, and a sharp rise with increasing periods to the break, followed by a gradual decline to longer periods. This is consistent with other analysis at the same time \citep{catshao,how12,dz13}. With the accumulation of additional data and more detailed treatment of selection effects, subsequent analyses favored a flatter distribution extending to smaller radii \citep{fress13,pet13b,sil15,tru16}, and a distribution falling off inversely with period ($\beta \sim -1$) at longer periods \citep{pet13a,sil15}. The plateau at small radii is also found around lower mass hosts \citep{dre13,dre15,muld15}.\

\citet{bur15} have presented an extensive discussion of planet occurrence using the Q1-Q16 Kepler sample. For their baseline model, they found corresponding values of $\alpha_1=-1.54 \pm 0.50$ and $\beta_2=-0.68 \pm 0.17$, with only weak evidence for a break in radius and assuming no break in period (they considered only periods $>50$ days and radii $<2.5r_{\earth}$). This is perhaps the most directly comparable to our analysis, as it uses the completeness estimates from \citet{chr15}; where this study uses the updated \citet{chr17} completeness data. We find very similar values  ($\alpha_1=-1.65\pm^{0.05}_{0.06}$; $\beta_2=-0.64\pm0.02$) in a comparable regime. In particular, we note that both of these studies find an increasing occurrence of small radius planets down to the detection threshold, a result also supported by another Bayesian methods estimate in \cite{hsu18}.\

Previous studies have used more limited parameter ranges to avoid issues of parameter covariance and susceptibility to completeness mapping. We approach the problem with a rigorous treatment of completeness mapping and a larger parameter space, recovering a similar power-law distribution. This congruity is an encouraging sign as it shows that the inclusion of a larger parameter space does not largely effect the model being inferred. Our inclusion of a broader range of periods and radius allow us to constrain the power-law uncertainty for radius and period to 3.8\% and 5.4\% respectively.\ 

We find breaks in our period and radius distributions occur at $p_{br}=7.08\pm^{0.32}_{0.31}$ days and $r_{br}=2.66\pm0.06 r_{\earth}$. These results are consistent with those found by prior authors.

\begin {table*}
\caption {A representation of the expected empirical multiplicity as a function of selection effects. Each column shows the expected population using the best fit model from this study (see Figure \ref{fig:post}). Starting from the left, moving right, each effect is adding in addition to all previous effects. The Multiple Detection Efficiency is broken into two columns. The Data column directly used the multiplicity values shown in Figure \ref{fig:surv}. In contrast, the Model column uses the modified Poisson distribution inferred from the multiplicity data ($\lambda=8.40 \pm 0.31$ and $\kappa=0.70 \pm 0.01$). \label{tab:dich}} 
\begin{tabular*}
{1 \textwidth}{@{\extracolsep{\fill}}l c c c c c c c} 
\hline \hline \multicolumn{1}{l}{} & \multicolumn{1}{c}{Geometric} & \multicolumn{1}{c}{Mutual Inclination} & \multicolumn{1}{c}{Single Detection} & \multicolumn{1}{c}{Multiple Detection}& \multicolumn{1}{c}{Multiple Detection}& \multicolumn{1}{c}{Real \emph{Kepler}}\\
\multicolumn{1}{l}{} & \multicolumn{1}{c}{} & \multicolumn{1}{c}{} & \multicolumn{1}{c}{Efficiency} & \multicolumn{1}{c}{Efficiency (Data)}& \multicolumn{1}{c}{Efficiency (Model)}& \multicolumn{1}{c}{Data}\\
\hline 
$Singles$    & $1870$ & $1910$ & $1558$   & $1649 \pm 71$ &  $1629\pm 61$ & $1637$\\
$Doubles$    & $686$  & $816$    & $397$    & $374 \pm 29$    &  $375 \pm 33$    & $346$\\
$Triples$    & $354$  & $483$    & $115$    & $103 \pm 15$    &  $113 \pm 15$  & $119$\\
$Quadruples$ & $207$   & $282$  & $30$   & $26 \pm 6$      &  $25 \pm 6$     & $43$\\
$Quintuples$ & $127$   & $159$   & $8$     & $5 \pm 3$     &  $5\pm 3$      & $13$\\
$Sextuples$  & $132$   & $77$     & $1$      & $1 \pm 1$      &  $1 \pm 1$    & $2$\\
$Septuples$  & $167$   & $28$      & $0$      & $0 \pm 1$      &  $0 \pm 1$    & $1$\\
\hline 
\end{tabular*}
\end{table*}

\subsection{Survival Function} 

\begin{figure}
\begin{center}
\includegraphics[width=8.5cm]{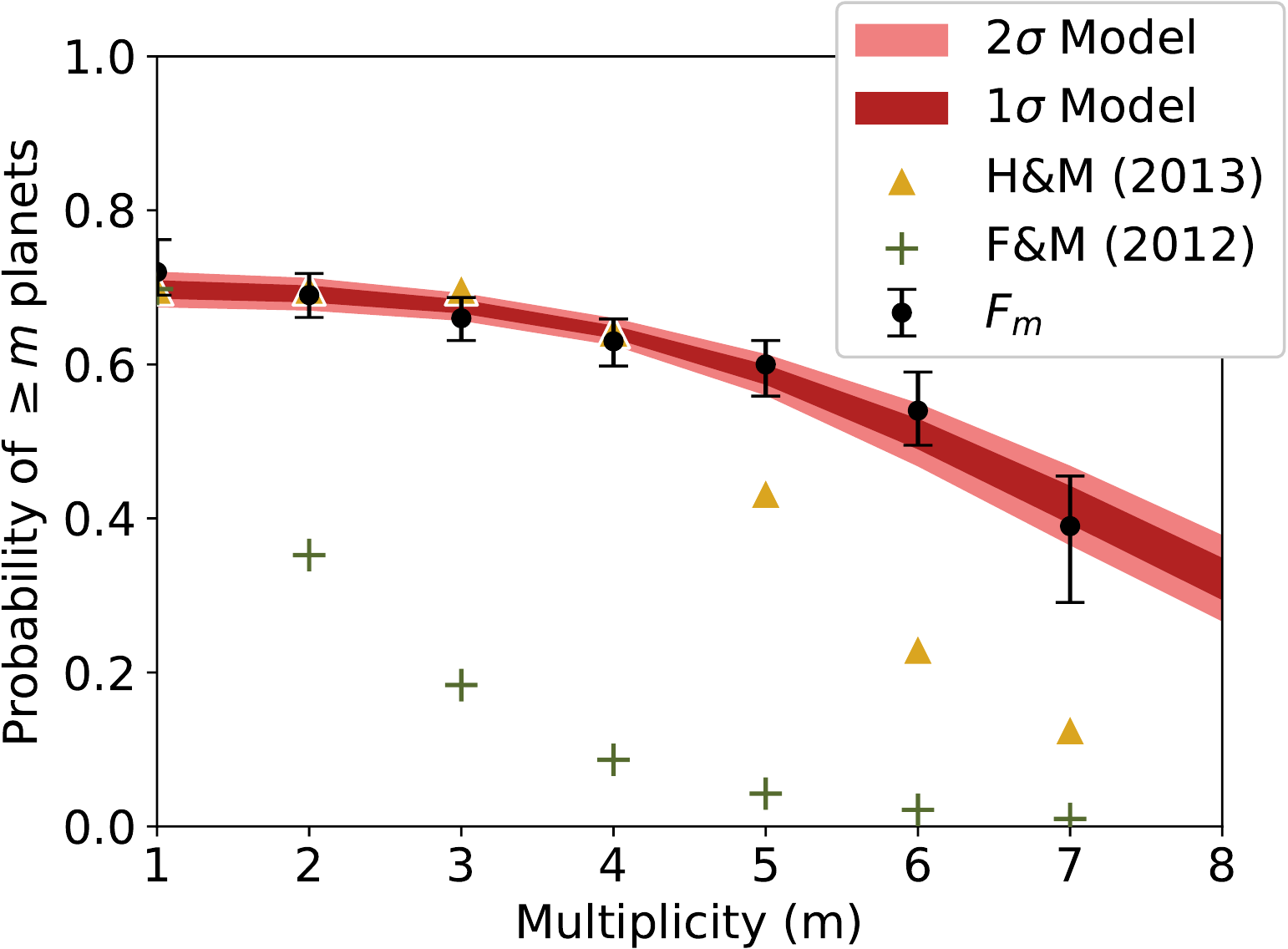}
\end{center}
\caption{A plot of the modified Poisson survival function and the system fraction provided by our Bayesian analysis. This model is fit using a likelihood maximization technique, with the assumption of Gaussian uncertainty (essentially, a $\chi^2$ minimization). The posterior distribution for the model is plotted by sampling 5000 models from the parameter posterior distributions in red. The dark red represents the $1\sigma$ range and the light red indicates the extent of the $2\sigma$ range. We have included the models provided by \citet{han13} (gold $\triangle$ makers) and \citet{fan12} (olive green + markers) for comparison. Both \citet{han13} and \citet{fan12} models have been renormalized by our best fit $\kappa$ value. \label{fig:surv}}
\end{figure}

Within this study, we only use planets provided by the \emph{Kepler} pipeline. The highest multiplicity seen is m=7 for a GK type star. This is certainly not the actual highest multiplicity within this parameter space. \citet{sha18} use a deep convolutional neural network to extract an 8th planet from the \emph{Kepler}-90 light curve, proving this assertion to be true. Using a Poisson survival function we can extrapolate the probability of existence for these higher multiplicity systems. The $F_m$ values found by this study represent the fraction of stars with at minimum $m$ planets. This lends itself well to a survival function, where the probability of existing up to a certain value (or multiplicity) is obtained. Survival functions ($S(x)$) can simply be written as: 

\begin{equation}
S(x)=1-CDF(x) 
\end{equation}

where $CDF(x)$ is the cumulative distribution function of the model. In this case we use a modified Poisson distribution to model multiplicity. Poisson distributions are ideal for planet multiplicity as these distributions are used for counting statistics. The modification is that the distribution is not truly normalized, but rather some fraction $\kappa$ of one. Now that that distribution is no longer normalized the survival function must be modified slightly ($S(x)=\kappa-CDF(x)$). This modification allows for an excess or scarcity of zero planet systems. We are only interested in stars that do harbor planets, thus this modification is necessary. The CDF for this modified function is given as:

\begin{equation}
CDF(m)=\sum_{n=1}^{m}\kappa\frac{\lambda^{n}e^{-\lambda}}{(n)!} \label{equ:cdfp}
\end{equation}

where $\kappa$ and $ \lambda$ are both fit parameters. Further discussion of this modified Poisson distribution can be found in Section 2.3 of \citet{fan12}. The results of this fit are presented in Figure \ref{fig:surv}. We find that $\lambda=8.40 \pm 0.31$ and $\kappa=0.70 \pm 0.01$ provide the best match for this distribution. This large $\lambda$ value incorporates a non-negligible fraction of systems with $m>10$. Since Equation \ref{equ:cdfp} allows for an inflated number of star without planet, the global average for GK dwarfs in the \emph{Kepler} parameter space (denoted as $\langle  N_{pl} \rangle $) can be found by multiplying $\lambda$, an estimate of the average number of planets a planet harboring system will contain, by $\kappa$, the fraction of stars that do harbor planets. We find that $\langle  N_{pl} \rangle =5.86 \pm 0.18$ planets per star. This is likely a lower bound as we have excluded the single Jupiter sized planets that have cleared their systems through migration. Since these stars are currently assumed to have zero planets by this paper, inclusion of these additional planets would increase the $\kappa$ value. However, we would expect our $\lambda$ parameter to slightly decrease, with the inclusion of these additional singles, as this value only considers systems that do harbor planets. Overall the increase in $\kappa$ will dominate, leading to an overall increase in $\langle  N_{pl} \rangle $.\

Previous studies have averaged over multiplicity and inferred the $\langle  N_{pl} \rangle $ value alone. These values are more difficult to compare as they are strongly dependent on the range of planet radius and period include in each study. Looking at short period ($p<50$ days) planets, \citet{you11} found $\langle  N_{pl} \rangle =1.36$. Using our population parameters and making similar cuts we find a comparable value ($\langle  N_{pl} \rangle =1.34\pm.06$). Turning the focus towards small planets ($.75<r<2.5r_{\earth}$) and long periods ($50<p<300$ days), \citet{bur15} found $\langle  N_{pl} \rangle =0.73\pm^{.19}_{.07}$. When we apply these same bounds to our model we again find a slightly larger value ($\langle  N_{pl} \rangle =1.15\pm.03$). The most comparable parameter space to our study is that of \citet{tru16}, who finds $\langle  N_{pl} \rangle =5.04\pm.23$ using a nearly identical parameter range. While there appears to be a mild tension with this value, we note that \citet{tru16} includes a much broader stellar temperature range and pre-\emph{Gaia} radius measurements, likely leading to this deflated $\langle  N_{pl} \rangle $ value. \  

With this function in hand, we can extrapolate to higher multiplicity. For example, our model suggests that $32.3 \pm 2.7\%$ of GK stars will harbor at least 8 planets within the \emph{Kepler} parameter space. In the parameter space of the \emph{Kepler} survey, our solar system has two planets (Venus and Earth). The radius of Mercury is slightly smaller ($.387 r_{\earth}$) than our range allows. Since we find $\langle  N_{pl} \rangle =5.86 \pm 0.18$ planets per solar-like star in this range, it appears that our system is more underpopulated than most other systems within $p<500$ days. We would expect $30\pm1\%$ systems harbor zero planets, $4.0\pm4.6\%$ harbor just one planet, and $2.0\pm4.2\%$ harbor only two planets within the range of this study. This lack of multiplicity in our solar system could be important for habitability, but such claims still lack strong evidence.\

\subsection{\emph{Kepler} Dichotomy} 

Analysis of the statistics of the \emph{Kepler} multiple planet systems \citep{lis11,fan12,han13,bal16} suggest that the underlying planetary population requires a two component model. One component is composed of systems with high planet multiplicity and a low inclination dispersion, while the other requires either low intrinsic multiplicity or a large inclination dispersion to reduce the frequency of transits by multiple planets. This has been termed the \emph{Kepler} dichotomy. \citet{lis11} inferred that the two populations had roughly equal frequencies and subsequent analyses confirmed this. There have been several models proposed to explain this on dynamical grounds \citep{joh12,mor16,han17}. The simplest solution is to consider a single population of planets in which some fraction have experienced excitation of their mutual inclinations. However, to meet the requirements of the transit statistics, the excitation is sufficiently large that dynamical stability is hard to maintain \citep{han17}. Thus, the \emph{Kepler} results seem to imply the existence of a low multiplicity population of planetary systems, whether due to formation or later dynamical instability.\

However, this finding rests on the relative frequencies of systems with single transiting planets versus multiple transiting planets. If the completeness is a function of the detection order, this may weaken the claim for a \emph{Kepler} dichotomy. In Figure \ref{fig:surv} we show that a single Poisson distribution can account for the multiplicity probabilities ($F_m$) extracted from our analysis. We find a much smaller fraction of intrinsically single systems than \cite{fan12} and find a distribution broadly similar to the model for a single, dynamically motivated population described in \cite{han13}. However, we still find $\sim 6\%$ of stars harbor intrinsically single or double planet systems. To test the robustness of this low multiplicity contribution we forward model the inferred population using the Poisson multiplicity model. In Table \ref{tab:dich} under the label ``Multiple Detection Efficiency (Model)'' we present the multiplicity results of this model. We can see that almost all of the empirical population fall within $1\sigma$ of the multiplicity model. This indicated that that apparent deviations in our infer $F_m$ values can be described by statistical fluctuations in population. Additionally, our $F_m$ are very dependent on the choice of mixture values displayed in Table \ref{tab:entag}. A proper accounting of these values would require distribution dependence. Averaging over these parameters, as done here, can cause mild deviations in the inferred $F_m$ values.\

In extracting the population $F_m$ values, we have only employed a mild Rayleigh distribution to account for mutual inclination of each system as directed by \citet{fan12} and have no larger inclination component. It appears that accounting for systematic loss of planets at higher multiplicity substantially reduces the low multiplicity population inferred as per the \emph{Kepler} Dichotomy. We shall now discuss how this works. \     

\begin{figure*}
\hspace*{-1.5cm}
\includegraphics[width=21cm]{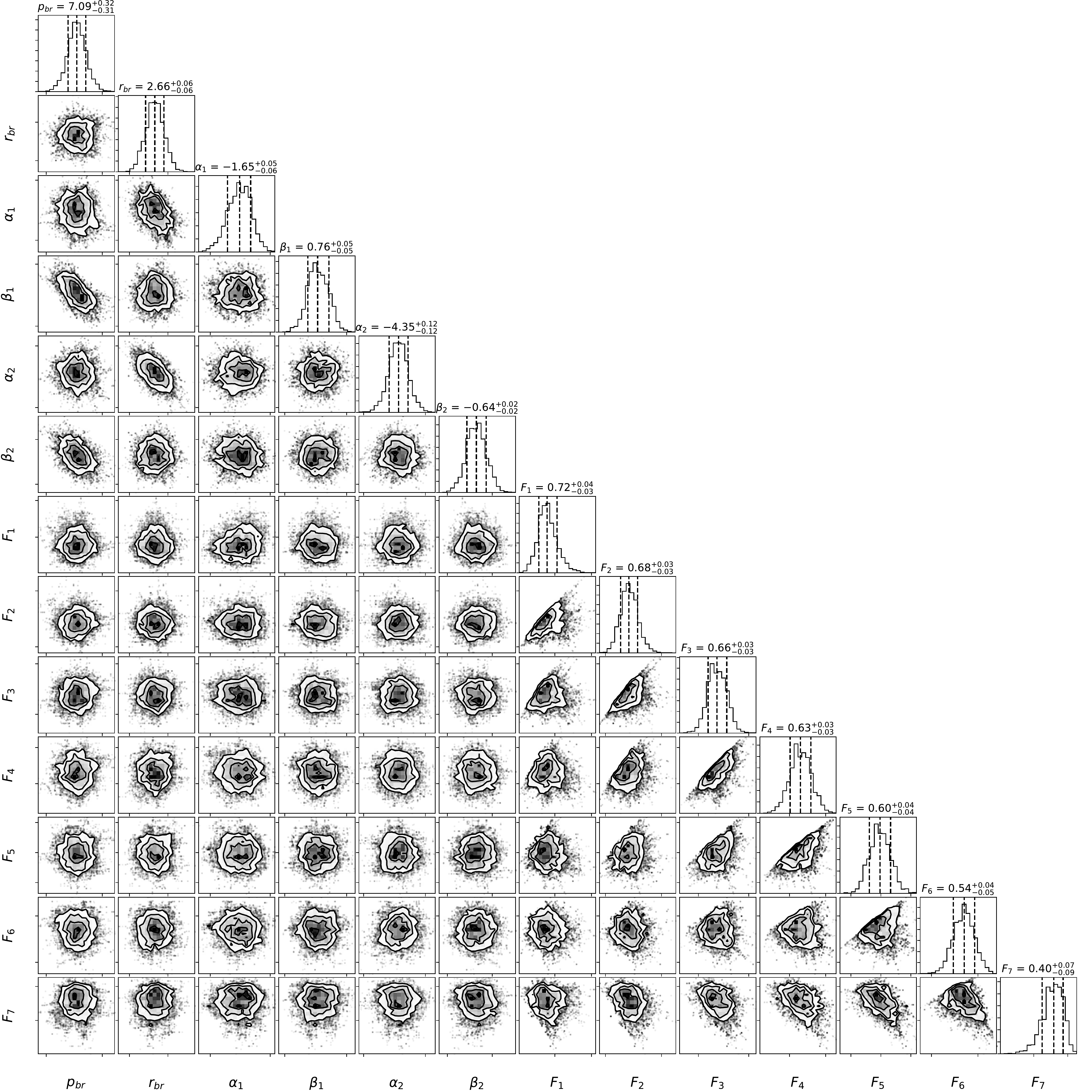}
\caption{The posterior distributions for the 13 parameters varied in this study. This is achieved using a burn-in of 100,000 steps and 20,000 steps to sample the posterior. The results of the fit are presented above the marginalized distribution of each parameter. The uncertainty is presented with a 68.3\% confidence interval. \label{fig:post}}
\end{figure*}

Using the forward model presented in Section \ref{sec:mult}, we look at how the inclusion of detection efficiency affected the gap seen between systems with one transiting planet and those with two transiting planets. The population provided by the parameters in Figure \ref{fig:post} is modeled 20 times and the median from each group is recorded in Table \ref{tab:dich}. Using our population parameters and a mild mutual inclination model show that this anomaly is largely due to \emph{Kepler} detection efficiency. Table \ref{tab:dich} shows how the frequency of detected systems of different transit multiplicity changes as we include different systematic effects. In the first column, we include only the correction of the probability of transit due to geometric alignment. For a simple numerical comparison, this results in a ratio of double transit to single transit systems of $0.37$, to be compared to the observed value of $0.21$ (the rightmost column). The inclusion of a small mutual inclination dispersion, comparable to that of \citet{fan12}, does not improve the ratio (second column). In the third column, we show the model in which we include the completeness corrections from \citet{chr17} without the multiplicity treatment discussed here. This results in a partial improvement of the ratio to $0.25$. It is also notable that the number of expected high transit multiplicity systems also drops significantly with the inclusion of this effect. Finally, in the fourth and fifth column, we show the expected numbers including the full, multiplicity-dependent completeness correction discussed here (Section \ref{sec:probd2}). We find that the expected number of different transit multiplicities are now very well matched to the observed numbers, substantially weakening the need for an additional population to explain the observations.\

The ultimate reason for this is that high transit multiplicity systems usually contain several planets that lie in the low MES region of parameter space, so that the incompleteness (especially when including the detection order effects) knocks planets down the multiplicity scale, resulting in many single transit systems that, in an ideal world, would show two or three transiting planets. Furthermore, the improved stellar radius measurements from \emph{Gaia} suggests that many stars have larger radii than previously believed \citep{ber18}. Increasing the stellar radius of system will decreases the probability of detection for an exoplanet. This correction will overall increase the inferred occurrence measurements.\

It is important to remember that our dataset does not include single hot Jupiter planets as discussed in Section \ref{sec:plan}. This observed population of 120 planets does not follow our power-law trend and appears to be uniquely single \citep{stef12}. While these outliers do provide some type of population dichotomy, their presence is not the most prominent cause of the excess of singles.\  
 
Our extracted population parameters $F_1$ and $F_2$ indicate that $4.0\pm4.6\%$ of the underlying population does have only one planet, and that this contribution can be described by the modified Poisson distribution used to fit the higher multiplicity systems. There is dynamical evidence that single transiting systems are more dynamically excited than multiple systems \citep{mort14,xie16,vanE18b} and this is consistent with the notion that some fraction of compact planetary systems are dynamically perturbed by the existence of giant planets on larger scales. Previously, \citet{han17} found that explaining the original excess of single transits required a frequency of giant planets on large scales that was roughly double that found by radial velocity surveys. The reduction found here substantially alleviates that discrepancy.\

Other recent work also supports the notion that single transiting systems are drawn from the same underlying planetary population as multiple harbor systems. \citet{weiss18} find that both populations share essentially the same stellar and planetary properties, while \citet{zhu18} use transit timing variations to infer that there is a strong correlation between multiplicity and dynamical excitation. They reject the notion that this is driven by giant planet excitation because they see no correlation with the metallicity of the host star, but such a correlation would be difficult to see at the level of $4\%$ as found here. This is further supported by \citet{MRK18}, who find no metallicity difference between hosts of single and multiple transiting systems, but could easily accommodate mixtures at the 50\% level.

\subsection{Considering Eccentricity}
\label{sec:ecc}

\begin{figure}
\begin{center}
\includegraphics[width=8.5cm]{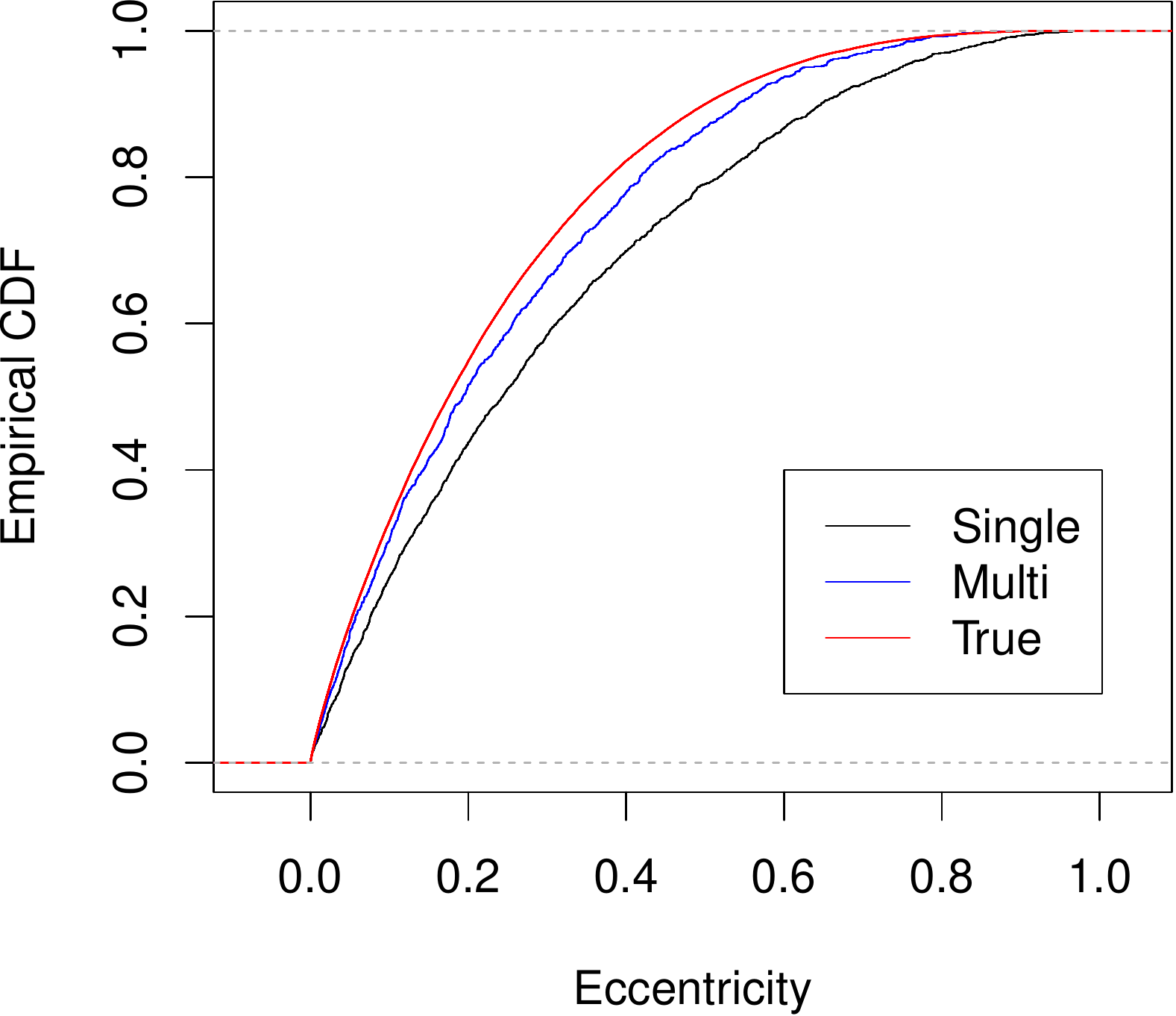}
\end{center}
\caption{A CDF showing the retrieved eccentricities from our forward modeling pipeline. The red line illustrates the eccentricities used to draw the underlying the Beta distribution \citep{kip13}. The black line represents the empirical CDF of the detected single planet systems and the blue line represents the eccentricities of the detected multi-planet systems. \label{fig:ecc}}
\end{figure}

Including eccentricity into our model increases the number of detected planets. We find that the best fit multiplicity parameters are as follows: $F_1=0.72\pm0.05$, $F_2=0.66\pm0.03$, $F_3=0.63\pm0.03$, $F_4=0.60\pm0.03$, $F_5=0.56\pm0.03$, $F_6=0.51\pm0.04$, and $F_7=0.43\pm0.07$. These parameters are fit using an analog to the \citet{han13} eccentricity model. The original modified Gamma distribution (scale=0.055) is unique to \citet{han13}. We map this model to a Beta distribution ($a=1.80$ and $b=14.46$), widely used among recent authors, for consistency. This model was inferred by simulating in situ gravitational assembly of planetary embryos and observing the resulting eccentricity population of the fully formed planets. Although derived within a specific scenario, this distribution matches well with a model in which planets explore the full range of available phase space subject to the constraint of dynamical stability \citep{tre15}. As such, it represents a plausible description of the level of eccentricity to be expected in such systems. The average eccentricity of this population is $\langle  e \rangle =0.11$. Comparing these values to those of our base model, we find that eccentricity flattens the CDF of planet multiplicity, slightly decreasing $\langle  N_{pl} \rangle $ to $5.69\pm0.17$ planets.\ 

Recently, \citet{vanE18b} provided evidence for two distinct populations of eccentricity (multi-planet systems and single planet systems). Using our forward modeling software (\textbf{\href{https://github.com/jonzink/ExoMult}{ExoMult}}), we test the strength of this hypothesis. Implementing only one true underlying eccentricity model, we inspect the detected eccentricity populations from both the single and multi-planet populations. When tested with the \citet{han13} model ($\langle  e \rangle =0.11$), we find no significant difference between the the observed eccentricities of multi-planet and single planet systems. This indicates that the differences noted by \citet{vanE18b} may be real. However, \citet{vanE18b} suggests a Beta distribution for single planet systems with $\langle  e \rangle =0.26$. This is a significantly larger average eccentricity than expected by the \citet{han13} model.\

When larger eccentricities are tested we do find observable differences between the single and multi-planet systems. The \citet{kip13} model ($a=0.867$ and $b=3.03$) was calculated using radial velocity discoveries and contains a significant fraction of massive planets. This distribution is probably too eccentric ($\langle  e \rangle =0.22$) for the tightly packed model discussed here, but illustrates the effects of detection bias on the eccentricity population. In Figure \ref{fig:ecc} we present the results of our test on the \citet{kip13} model. We find that multi-planet systems tend to produce more low eccentricity detections than single planet detections despite being drawn from the same underlying population. Analyzing the statistical difference with an \emph{Anderson-Darling} test produces a P-value of $10^{-7}$, suggesting these differences would appear statistically significant. Furthermore, we can see that neither of the detected populations closely mimic the true Beta distribution, highlighting the importance of detection efficiency consideration when performing eccentricity occurrence measurements. This effect is caused by the increased transit duration for higher eccentricity transits. Increasing the transit duration improves the planet MES, making the signal easier to detect. Since the highest MES planets are the most likely to be detected, this biases the empirical population toward higher eccentricity. The sorting order in combination with the multiplicity detection efficiency of the \emph{Kepler} pipeline will further exaggerate this bias in the single planet systems.\

It is clear that low eccentricity distributions are less affected by this bias. Manually tuning the Beta distribution we find that models with $\langle  e \rangle \ge0.18$ will produce statistically significant (P-value$\le 0.001$) differences between the empirical eccentricity population of singles and multiple planet systems. Since \citet{vanE18b} suggests a $\langle  e \rangle =0.26$ model for the singles and a $\langle  e \rangle =0.05$ model for the multi-planet systems, it is difficult to determine the effect of detection bias on their eccentricity model. At this point we cannot rule out that two distinct populations of eccentricity exist between the single and multi-planet systems, but propose that such claims require further evidence.\

\subsection{Extrapolation to Longer Periods}

As mentioned above, our general populations parameters do not differ greatly from those of previous studies. 
 The quantity $\Gamma_{\earth}$ is often quoted to avoid any need for understanding the habitable zone or habitable radius range. 
\begin{equation}
\frac{dN}{d\textup{ln}p_{\earth}\textup{ }d\textup{ln}r_{\earth}}=\Gamma_{\earth}
\end{equation}

We find $\Gamma_{\earth}=1.31\pm0.07$, consistent with the previous value of \cite{bur15} ($\Gamma_{\earth}=0.6$ with a range of 0.04 to 11.5). \cite{you11} found a much higher value of $\Gamma_{\earth}=2.75\pm0.3$, when extrapolating from periods $<50$ days. The lack of long period planets provided a weaker power-law, producing the inflated $\Gamma_{\earth}$ value. Furthermore, we find tension with \cite{for14} (with $\Gamma_{\earth}=0.019\pm^{0.019}_{0.010}$). \citet{for14} avoid the assumption of a particular functional form for the extrapolation to longer periods, by using a Gaussian process regression to determine the shape of the distribution. However, they use the results of the \emph{TERRA} pipeline in it's original form, in which it only reported the highest signal to noise candidate around each star. Although they back out an estimate of the detection efficiency from the results of \citet{pet13b}, we have shown in Section \ref{sec:like} that detection order can bias the results. In particular, we expect \citet{for14} to undercount small planets and long period periods. Both of these biases will lower the $\Gamma_{\earth}$ value and we should regard the \citet{for14} result as a lower limit.\

For the occurrence of habitable planets we follow the procedure provided by \cite{bur15}. This $\zeta_{\earth}$ value is found by integrating the population distribution by 20\% of $r_{\earth}$ and $p_{\earth}$ in both directions. We find $\zeta_{\earth}=0.217\pm0.014$ using our inferred population parameters, similar to the $\zeta_{\earth}=0.10$ (with a range of 0.01 to 2) found in \cite{bur15}. \

\section{Conclusion}
\label{sec:conc}

We present a new method for determining the frequency of exoplanet multiplicity within the \emph{Kepler} dataset. In doing so we provide the following new fitting features and conclusions:\\

\noindent
1.\indent Previous studies have discussed and provided methods for calculating high multiplicity transit probabilities (\citet{rag10,bra16,rea17}). For occurrence calculations these procedures are often too complex and computationally expensive to carry out. We provide a new method which marginalizes over mutual inclination and the empirical \emph{Kepler} period set to determine the transit probabilities for \emph{Kepler} multi-planet systems. Using this, we provide the transit probabilities for multiple systems containing up to 7 planets. This simplification is important and useful when trying to fit multiplicity parameters via MCMC or some other fitting method that requires $10^4$ calculations.\

Our method does make some simplification assumptions in the interests of speed. We assume the measurements of planet radius and period are perfect. The uncertainty in period is negligible, however the radius measurements retain significant uncertainty and the present dispersion may yet mask finer features in the distribution. In accounting for mutual inclination, we adopt the model provided by \citet{fan12}. This is derived using a different multiplicity model than that found here. All orbits are assumed to be circular in our base model. Because many of the systems are very compact, circular orbits are required for any type of stability. Tidal circularization will also force many of these planets into circular orbits. However, it is possible that some portion of the population, investigated here, contains varying amounts of eccentricity. We show that any amount of eccentricity will increases our the overall multiplicity values, but decreases the fraction of systems with planets. We have assumed the appropriate model for exoplanet occurrence is a broken power-law. Furthermore, we assume period and radius and uncorrelated. It has been shown by \citet{owe13} and \citet{wei17} that a mild correlation exist between period and radius at short periods where photoevaporation can take effect. Nevertheless, the fact that our forward modeling matches the data inspires confidence that the model provides a coherent description of the data.\\

\noindent
2.\indent In systems with more than one detected planet, we find that detection efficiency decreases for higher detection order planets. This conclusion was achieved by re-visiting the \citet{chr17} injections and looking at systems with pre-existing planets. Multiple planets systems experience an additional loss,  for lower MES planets within each system, of at least 5.5\% and 15.9\% for periods $<200$ days and $>200$ days respectively. This type of increased selection effects indicates that a larger fraction of the population is being missed. Being able to infer a larger population of multiple exoplanet systems significantly decreases the gap between single and double planet systems. The initial motivation for additional detection efficiencies for multi-planet systems, was the 61 known KOIs lost during the \citet{chr17} injections. When testing our additional selection effects, for multiples, we expect $41\pm7$ planets should be lost due to a similar type of injection test. Because we find that 61 KOIs are lost (rather than 41) we suspect higher order detection efficiencies may be necessary for an accurate accounting of the true underlying populations.\\

\noindent
3.\indent Using Bayesian statistics, we expand the Poisson process likelihood to account for variations in detection order. Furthermore, we are able to infer population multiplicity from this fitting process. The results from this fit match that of \citet{bur15}, but provide an improved measurement with reduced uncertainty from \emph{Gaia}, CKS, and asteroseismology \citep{pet17,joh17,ber18,vanE18a}. Furthermore, by looking at the occurrence of single and double-planet systems, we only find a $0.9\sigma$ difference between these two populations ($4.0\pm4.6\%$). This disparity can be explained by a modified Poisson distribution with $\lambda=8.40 \pm 0.31$ and $\kappa=0.70 \pm 0.01$, indicating that the \emph{Kepler} Dichotomy (discussed by \cite{lis11,fan12,han13,bal16}) may largely be an artifact of detection efficiency and statistical fluctuation.\

Using a Poisson process likelihood requires that each planet is drawn independently, which is clearly not the case for planets in multiple systems. Much of the work in this study is accounting for these dependencies. Ignoring the independence requirement of Poisson process could be suspect, but is again justified by the success of our forward model, where this assumption is not necessary. The independence of radius between planets within a system has also not been accounted for within this study.\\ 

\noindent
4.\indent Given our inferred multiplicity model we can extrapolate to higher multi-planet systems. We find that $32.3 \pm 2.7 \%$ of solar-like stars should contain at least 8 planets within 500 days. The existence of a single 7 planet system and a single 8 planet system (Kepler 90) indicates these systems should be rare but still detectable. We would expect to find $<1$ eight planet systems within the constraints of this study.\\

\noindent
5.\indent We introduce (\textbf{\href{https://github.com/jonzink/ExoMult}{ExoMult}}) and demonstrate that forward modeling a broken power-law distribution can still provide a reasonable model for the exoplanet population, despite growing evidence for a gap in $1.5-2r_{\earth}$ range \citep{ful17,ber18,weiss18}. We find that our fitting model also produces similar populations of multiplicity to that of the empirical \emph{Kepler} data set, indicating the success of this method.\\

\noindent
6.\indent Using the the eccentricity model of \citet{han13}, we show that eccentricity can affect the multiplicity occurrence by slightly decreasing the expected number of planets around each star. We also find that for eccentricity models with $\langle  e \rangle \ge0.18$ the \emph{Kepler} pipeline will significantly skew the empirical population of eccentricity for single transiting systems, suggesting that differences seen between the single and multiple planet systems may be artificial.\\

\subsection{Future Goals} 
As mentioned previously, the uncertainties in the radius measurement are still quite large. Using a Bayesian hierarchical model, this uncertainty can be incorporated when fitting for population parameters (see \citealt{for14}). We hope to include this feature into our next generation of occurrence fitting.\

The multiplicity parameters derived here can be use in determining an Eta Earth measurement. The importance of neighboring planets could be essential for the long term stability of an Earth analog \citep{hor17}, thus it is important to understand the likelihood of this Earth analog within a multiple system.\

The new detection efficiency is limited to $m\geq2$. Ideally, we would want the detection efficiency for each detection order. To do so one would need to perform an alternative injection experiment, where numerous planets are injected into each system and the recovery of each order can be better sampled. It would also be useful to understand the effects of resonance on detection efficiency. Looking at a select group of stars and injecting many planets at various period ranges could provide an understanding of these features (as performed by \citealt{bur17b}).\

With the loss of \emph{Kepler} and the upcoming release of \emph{TESS} it will be essential to combine data across missions to calculate a more robust occurrence measurement. Doing so will require accounting for differing detection efficiencies across each mission. The method described here may provide a unique way of incorporating these different selection effects while producing a uniform population distribution.\    

\section*{Acknowledgement}
We would like to thank the anonymous referee for useful feedback. The simulations described here were performed on the UCLA Hoffman2 shared computing cluster and using the resources provided by the Bhaumik Institute. This research has made use of the NASA Exoplanet Archive, which is operated by the California Institute of Technology, under contract with the National Aeronautics and Space Administration under the Exoplanet Exploration Program.

\end{document}